\documentclass[prb,aps,twocolumn]{revtex4-1}
\usepackage{graphics, bm, psfrag, amsmath, amssymb, epsfig, grffile, float, bbold}
\usepackage{subfigure}
\usepackage{dsfont}
\usepackage{color}
\usepackage{hyperref}
\newcommand*{\citen}[1]{%
  \begingroup
    \romannumeral-`\x 
    \setcitestyle{numbers}%
    \cite{#1}%
  \endgroup   
}

\begin{document}

\title{Odd-frequency pairing and Kerr effect in the heavy-fermion superconductor UPt$_3$}

\author{Christopher Triola}
\affiliation{Department of Physics and Astronomy, Uppsala University, Box 516, S-751 20 Uppsala, Sweden}
\author{Annica M. Black-Schaffer}
\affiliation{Department of Physics and Astronomy, Uppsala University, Box 516, S-751 20 Uppsala, Sweden}

\begin{abstract}
We study the emergence of odd-frequency superconducting pairing in UPt$_3$. Starting from a tight-binding model accounting for the nonsymmorphic crystal symmetry of UPt$_3$ and assuming an order parameter in the $E_{2u}$ representation, we demonstrate that odd-frequency pairing arises very generally, as soon as inter-sublattice hopping or spin-orbit coupling is present. We also show that in the low temperature superconducting $B$ phase, the presence of a chiral order parameter together with spin-orbit coupling, leads to additional odd-frequency pair amplitudes not present in the $A$ or $C$ phases. Furthermore, we show that a finite Kerr rotation in the $B$ phase is only present if odd-$\omega$ pairing also exists.

\end{abstract}


\maketitle
\section{Introduction}

The heavy-fermion material UPt$_3$ has a truly unconventional superconducting phase diagram, possessing two zero-field superconducting phases, the $A$ phase and the $B$ phase, with critical temperatures $T_{c,+}\approx 550$ mK and $T_{c,-}\approx 500$ mK,\cite{stewart_prl_1984,sauls1994} respectively. Additionally, a third phase, the $C$ phase, emerges at high magnetic field.\cite{adenwalla1990} Knight shift observations point to a spin-triplet superconducting order parameter.\cite{tou_prl_1996} Josephson interferometry has revealed the presence of line nodes in the A phase,\cite{strand_science_2010} as well as the onset of a complex order parameter in the $B$ phase.\cite{strand_prl_2009,strand_science_2010} Recently, measurements of the Kerr effect have also demonstrated time-reversal symmetry breaking in the $B$ phase.\cite{schemm_2014} These observations appear to be consistent with a gap belonging to the $E_{2u}$ representation.\cite{sauls1994,strand_prl_2009,strand_science_2010,nomoto_prl_2016} Furthermore, recent work considering the nonsymmorphic crystal structure\cite{yanase_prb_2016,yanase_2017_prb} suggests that the measured value of the Kerr rotation is related to the presence of a nonunitary linear combination of $f$-wave and $d$-wave pairing in the $B$ phase.\cite{wang_2017} 

It is well-known that the fermionic nature of electrons tightly constrains the allowed symmetries of the superconducting gap function. Specifically, in the limit of equal-time pairing and a single-component gap, spatially even-parity gap functions (like $s$- or $d$-wave) must correspond to spin-singlet states, while odd-parity gap functions ($p$- or $f$-wave) must correspond to spin-triplet states. However, if the electrons comprising the condensate are paired at unequal times the superconducting gap can be odd in time or, equivalently, odd in frequency (odd-$\omega$). In that case the condensate can be even in spatial parity and  spin-triplet or odd-parity and spin-singlet. This possibility, originally posited for $^3$He by Berezinskii\cite{Berezinskii1974} and then later for superconductivity,\cite{kirkpatrick_1991_prl,belitz_1992_prb,BalatskyPRB1992} is intriguing both because of the unconventional symmetries which it permits and for the fact that it represents a class of hidden order, due to the vanishing of all equal time correlations. 

While the thermodynamic stability of intrinsically odd-$\omega$ phases is, so far, only discussed as a theoretical possibility,\cite{coleman_1993_prl,coleman_1994_prb,coleman_1995_prl,heid1995thermodynamic,belitz_1999_prb,solenov2009thermodynamical,kusunose2011puzzle,FominovPRB2015} significant progress has been made understanding systems with conventional superconductors in which odd-$\omega$ pairing can be induced by altering the native superconducting correlations.\cite{BergeretPRL2001, bergeret2005odd, yokoyama2007manifestation, houzet2008ferromagnetic, EschrigNat2008, LinderPRB2008, crepin2015odd, YokoyamaPRB2012, Black-SchafferPRB2012, Black-SchafferPRB2013, TriolaPRB2014, tanaka2007theory, TanakaPRB2007, LinderPRL2009, LinderPRB2010_2, TanakaJPSJ2012, parhizgar_2014_prb, triola2016prl, triolaprb2016} Well-established examples can be found in ferromagnet-superconductor junctions,\cite{BergeretPRL2001, bergeret2005odd, yokoyama2007manifestation, houzet2008ferromagnetic, EschrigNat2008, LinderPRB2008, crepin2015odd} in which experiments have recently observed key signatures of odd-$\omega$ pair correlations.\cite{di2015signature,di2015intrinsic} For a modern review of this field see Ref.~[\citen{linder2017odd}].

Another intriguing possibility for odd-$\omega$ superconductivity can be found in multiband superconductors in which it has been shown that odd-$\omega$ pairing is ubiquitously induced in the presence of interband hybridization.\cite{black2013odd,komendova2015experimentally,komendova2017odd,triola_prb_2017} As an illustrative example, if we consider a generic two-band model: $H=\xi_1\psi^\dagger_1\psi_1+\xi_2\psi^\dagger_2\psi_2+\Delta_1\psi^\dagger_1\psi^\dagger_1+\Delta_2\psi^\dagger_2\psi^\dagger_2+\text{H.c.}$, the addition of any finite hybridization term of the form $\Gamma\psi^\dagger_1\psi_2$ induces odd-$\omega$ interband pairing proportional to the difference of the two gaps, $\sim \omega\Gamma(\Delta_1-\Delta_2)$.\cite{black2013odd,triola_prb_2017} An interband hybridization of this form is intrinsic to the superconductor whenever there is a mismatch between the orbital character of the Cooper pairs and that of the quasiparticles of the normal state or, alternatively, it can arise from scattering processes in the presence of disorder.\cite{komendova2015experimentally,komendova2017odd} In contrast to other known mechanisms  for realizing odd-$\omega$ pairing which employ superconductor heterostructures, multiband superconductors allow for the generation of odd-$\omega$ pair amplitudes in the bulk, without breaking either spatial translation or time-reversal symmetry. With many known multiband superconductors with highly unconventional features, such as Sr$_2$RuO$_4$,\cite{maeno1994superconductivity,maeno2012} iron-based superconductors,\cite{hunte2008two,kamihara2008iron,ishida2009extent,cvetkovic2009multiband,kamihara2008iron,stewart2011superconductivity} MgB$_2$, \cite{nagamatsu2001superconductivity,bouquet2001specific,brinkman2002multiband,golubov2002specific,iavarone2002two} and UPt$_3$,\cite{stewart_prl_1984,adenwalla1990,sauls1994,strand_prl_2009,strand_science_2010} it remains a very interesting question how much odd-$\omega$ superconductivity contributes to their physical properties.

It was recently shown that the multiband superconductor Sr$_2$RuO$_4$ hosts odd-$\omega$ pairing due to the finite hybridization between the different orbitals in the normal state.\cite{komendova2017odd} At the same time it was shown that the conditions leading to the observation of a finite Kerr rotation angle also guarantee the emergence of odd-$\omega$ pairing. Since the Kerr effect has been observed in Sr$_2$RuO$_4$,\cite{xia_prl_2006} this directly implies that Sr$_2$RuO$_4$ hosts odd-$\omega$ pair amplitudes. The Kerr effect has also been observed in the multiband superconductor UPt$_3$;\cite{schemm_2014} however, while Sr$_2$RuO$_4$ is a relatively simple system assumed to possess a $p$-wave order parameter, the gap structure in UPt$_3$ is thought to be primarily $f$-wave and likely with additional character. Given that the Kerr rotation angle is known to be highly sensitive to system-specific details, such as the existence of interband transitions\cite{wysokinski_2012_prl,taylor_prl_2012,gradhand_2013_prb} or the presence of disorder,\cite{lutchyn_prb_2009} the observation of a finite Kerr rotation angle in UPt$_3$ cannot simply be presumed to imply the presence of odd-$\omega$ pairing also in this material. The purpose of this work is therefore to elucidate the possibility of odd-$\omega$ pairing channels in UPt$_3$ and, if possible, connect these to the Kerr effect found in the $B$ phase of UPt$_3$.  

In this work we use a tight-binding model, which, while simple enough to allow for analytical results, captures the $5f$ states on the U atoms, the main Fermi surfaces, and explicitly takes into account the nonsymmorphic symmetry of the lattice. Adhering to the bulk of experimental results we further assume spin-triplet pairing within the $E_{2u}$ irreducible representation. By calculating the full anomalous Green's function we are able to extract all possible odd-$\omega$ pairing amplitudes in UPt$_3$. We find that as soon as inter-sublattice hopping is present, i.e.~out-of-plane nearest neighbor U-U hopping, intra-sublattice odd-$\omega$ pairing emerges in all three superconducting phases of UPt$_3$. For finite spin-orbit coupling, we find that inter-sublattice odd-$\omega$ pairing is also always present in all three phases. In the $B$ phase we find additional inter- and intra-sublattice odd-$\omega$ pairing due to spin-orbit coupling and the nonunitary order parameter.
We furthermore compare the criteria for the existence of odd-$\omega$ pairing and finite Kerr effect, as experimentally measured in the $B$ phase of UPt$_3$. We are able show that the conditions needed for a finite Kerr rotation angle automatically lead to odd-frequency pairing. Thus, the finite Kerr effect measured in the $B$ phase serves as experimental evidence for odd-$\omega$ superconductivity in UPt$_3$.

The remainder of this article is organized as follows. In Section \ref{sec:model} we introduce the model used to describe the electronic properties of UPt$_3$ and discuss the assumptions used to solve it analytically. In Section \ref{sec:pairsym} we perform our analysis of the symmetries exhibited by the anomalous Green's functions, finding the conditions under which odd-$\omega$ pairing is expected to emerge in UPt$_3$. In Section \ref{sec:kerr} we compare these conditions to previous theoretical work characterizing the Kerr effect in UPt$_3$. In Section \ref{sec:con} we conclude our study.

\begin{figure}
 \begin{center}
  \centering
  \includegraphics[width=0.4\textwidth]{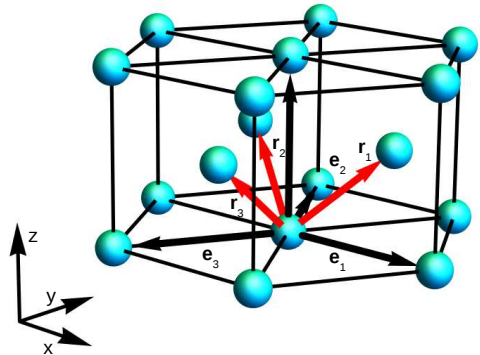}
  \caption{Schematic depiction of the locations of the Uranium atoms in the AB-stacked hexagonal lattice of UPt$_3$ with vectors in the basal plane labeled $\textbf{e}_i$ and inter-sublattice vectors labeled $\textbf{r}_i$.  }
  \label{fig:lattice}
 \end{center}
\end{figure}

\section{Model}
\label{sec:model}
In Fig.~\ref{fig:lattice} we show the three-dimensional (3D) crystal structure of UPt$_3$ with the U ions (blue spheres) forming AB-stacked layered triangular lattices with basal plane lattice vectors given by: $\textbf{e}_1=(1,0,0)$, $\textbf{e}_2=(-\tfrac{1}{2},\tfrac{\sqrt{3}}{2},0)$, $\textbf{e}_3=(-\tfrac{1}{2},-\tfrac{\sqrt{3}}{2},0)$ and inter-sublattice vectors given by: $\textbf{r}_1=(\tfrac{1}{2},\tfrac{1}{2\sqrt{3}},\tfrac{1}{2})$, $\textbf{r}_2=(-\tfrac{1}{2},\tfrac{1}{2\sqrt{3}},\tfrac{1}{2})$, $\textbf{r}_3=(0,-\tfrac{1}{\sqrt{3}},\tfrac{1}{2})$. The Pt ions (omitted for simplicity) are located between each of the nearest-neighbor U ions. To model the electronic properties of UPt$_3$ near the Fermi level we focus on the itinerant 5$f$ electrons originating from the U atoms. Motivated by previous work we assume the same tight-binding Bogoliubov-de Gennes Hamiltonian:\cite{yanase_prb_2016,yanase_2017_prb,wang_2017}
\begin{equation}
\begin{aligned}
\mathcal{H}&=\sum_{\textbf{k},m,\sigma}\xi_\textbf{k} c^\dagger_{\textbf{k}m\sigma}c_{\textbf{k}m\sigma} + \sum_{\textbf{k},\sigma}\left[\epsilon_\textbf{k}c^\dagger_{\textbf{k}1\sigma}c_{\textbf{k}2\sigma} + \text{H.c.}\right] \\
&+\sum_{\textbf{k},m,m',\sigma,\sigma'}g_\textbf{k} \tau^z_{mm'}\otimes\sigma^z_{\sigma\sigma'} c^\dagger_{\textbf{k}m\sigma}c_{\textbf{k}m'\sigma'} \\
&+\frac{1}{2}\sum_{\textbf{k},m,m',\sigma,\sigma'}\left[\Delta_{mm',\sigma\sigma'}(\textbf{k}) c^\dagger_{\textbf{k}m\sigma}c^\dagger_{-\textbf{k}m'\sigma'} + \text{H.c.}\right]
\end{aligned}
\label{eq:hamiltonian}
\end{equation}
where $c^\dagger_{\textbf{k}m\sigma}$ ($c_{\textbf{k}m\sigma}$) creates (annihilates) a fermionic quasiparticle with crystal momentum $\textbf{k}$, on sublattice $m=\{1,2\}$, and with spin $\sigma=\{\uparrow,\downarrow\}$. The intra-sublattice hopping terms are given by $\xi_\textbf{k}=2t\sum_i\cos\textbf{k}_{||}\cdot\textbf{e}_i + 2t_z\cos k_z-\mu$, with $\textbf{k}_{||}\equiv(k_x,k_y,0)$, which is manifestly even in $\textbf{k}$. The inter-sublattice hopping terms are given by $\epsilon_\textbf{k}=2t'\cos\tfrac{k_z}{2}\sum_ie^{i\textbf{k}_{||}\cdot\textbf{r}_i}$, which is, in general, complex with $\Re e\{\epsilon_\textbf{k}\}$ even in $\textbf{k}$, while $\Im m\{\epsilon_\textbf{k}\}$ is odd in $\textbf{k}$. The Kane-Mele-like spin-orbit coupling is described by $g_\textbf{k}=\alpha\sum_i\sin\textbf{k}_{||}\cdot\textbf{e}_i$ which is clearly odd in $\textbf{k}$. Following the same conventions, a general superconducting order parameter is written as $\Delta_{mm',\sigma\sigma'}(\textbf{k})$. As is widely done, we assume an order parameter within the  $E_{2u}$ irreducible representation and with spin-triplet $m_z = 0$ pairing.\cite{sauls1994,joynt_rmp_2002,yanase_prb_2016}

\begin{figure}
 \begin{center}
  \centering
  \includegraphics[width=0.4\textwidth]{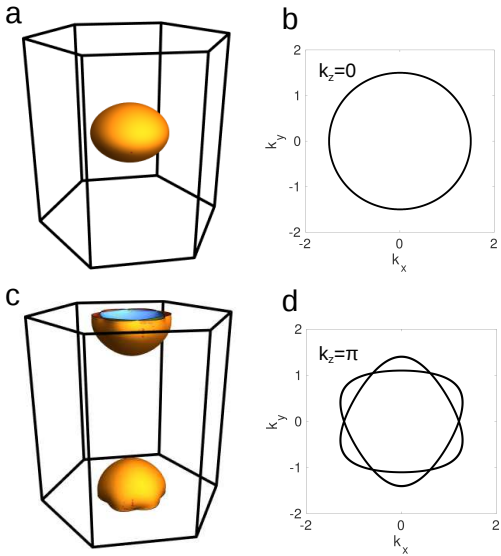}
  \caption{(a) Fermi surface of UPt$_3$ shown within the 3D Brillouin zone (hexagonal frame) plotted using the Hamiltonian in Eq.~(\ref{eq:hamiltonian}) with parameters $\left(t,t_z,t',\alpha,\mu\right)=\left(1,4,1,0,16\right)$ to reproduce the topology of the Fermi surface near the $\Gamma$-point. (b) A 2D cross-section of the Fermi surface shown in (a) for $k_z=0$. (c) Same as (a) except using different parameters, $\left(t,t_z,t',\alpha,\mu\right)=\left(1,-4,1,2,12\right)$, to reproduce the topology of the Fermi surface at the $A$-point. (d) A 2D cross-section of the Fermi surface shown in (c) for $k_z=\pi$. }
  \label{fig:fermi}
 \end{center}
\end{figure}

Quantum oscillation measurements employing the de Haas-van Alphen effect\cite{taillefer_1987,taillefer_prl_1988} combined with first-principles calculations\cite{wang_prb_1987,norman_ssc_1988,joynt_rmp_2002,mcmullan2008fermi,nomoto_prl_2016} have revealed several Fermi surfaces in UPt$_3$: two Fermi surfaces at the $A$-point, the so-called ``starfish" and ``octopus"; three Fermi surfaces at the $\Gamma$-point, the ``oyster", ``mussel", and ``pearl"; and also relatively small Fermi surfaces at the $K$-points, the ``urchins". The Hamiltonian in Eq.~(\ref{eq:hamiltonian}) can reproduce the topology of the Fermi surfaces appearing at the $\Gamma$-point using the parameters:\cite{yanase_prb_2016} $\left(t,t_z,t',\alpha,\mu\right)=\left(1,4,1,0,16\right)$, see Figs.~\ref{fig:fermi}(a, b). This same model can also reproduce the topology of the ``starfish" Fermi surface appearing at the $A$-point, using a different set of parameters:\cite{yanase_prb_2016} $\left(t,t_z,t',\alpha,\mu\right)=\left(1,-4,1,2,12\right)$, see Figs.~\ref{fig:fermi}(c, d). Therefore, the inter-sublattice hybridization as described by the Hamiltonian in Eq.~(\ref{eq:hamiltonian}) is necessarily appreciable at both $\Gamma$ and $A$, while the spin-orbit coupling is only relevant at $A$ due to its $\bf{k}$-dependence. In what follows, we derive general results without assuming a particular set of values for these parameters and proceed to discuss the implications considering the sets of parameters associated with the Fermi surfaces appearing at $A$ and $\Gamma$ separately. 

In general, a superconducting order parameter belonging to the $E_{2u}$ irreducible representation, as widely assumed for UPt$_3$, may be written as $\hat{\Delta}(\textbf{k})=\eta_1\hat{\Gamma}_1(\textbf{k})+\eta_2\hat{\Gamma}_2(\textbf{k})$ where $\hat{\Gamma}_i(\textbf{k})$ are basis functions and $\eta_i$ are complex numbers parameterizing the phase diagram.\cite{sauls1994,joynt_rmp_2002,nomoto_prl_2016,yanase_prb_2016} Previous analyses using Ginzburg-Landau theory have demonstrated that the minimum energy solution may be parameterized by a single real number, $\eta$, such that $(\eta_1,\eta_2)=\Delta_0 (1,i\eta)/\sqrt{1+\eta^2}$, where in the $A$ phase $\eta=\infty$, in the $B$ phase $0<\eta<\infty$, and in the $C$ phase $\eta=0$.\cite{sauls1994,joynt_rmp_2002,yanase_prb_2016} Following recent work,\cite{nomoto_prl_2016,yanase_prb_2016,wang_2017} we use basis functions that explicitly account for the symmetries of the lattice, which have been shown\cite{yanase_prb_2016} to give rise to a linear combination of $p$-wave, $d$-wave, and $f$-wave symmetries: 
\begin{equation}
\begin{aligned}
\hat{\Gamma}_1(\textbf{k})&=\left[\delta\left\{p_x(\textbf{k})\sigma^x\otimes\tau^0-p_y(\textbf{k})\sigma^y\otimes\tau^0 \right\} \right. \\
&+\left. f_{(x^2-y^2)z}(\textbf{k})\sigma^z\otimes\tau^x -d_{yz}(\textbf{k})\sigma^z\otimes\tau^y \right]i\sigma^y \\
\hat{\Gamma}_2(\textbf{k})&=\left[\delta\left\{p_y(\textbf{k})\sigma^x\otimes\tau^0+p_x(\textbf{k})\sigma^y\otimes\tau^0 \right\} \right. \\
&+\left. f_{xyz}(\textbf{k})\sigma^z\otimes\tau^x -d_{xz}(\textbf{k})\sigma^z\otimes\tau^y \right]i\sigma^y, 
\end{aligned} 
\label{eq:gap_basis}
\end{equation}
where $\delta$ is a small parameter associated with a weak $p$-wave component,\cite{yanase_prb_2016} while $\sigma^i$ and $\tau^i$ are Pauli matrices in spin and sublattice space, respectively. Notice the unusual combination of spin-triplet $f$-wave terms being odd in spatial parity and spin-triplet $d$-wave terms being even in parity. This combination is caused by the nonsymmorphic lattice symmetry. Note that these terms still satisfy the constraints imposed by Fermi-Dirac statistics on the Cooper pairs since the $f$-wave terms are even in the sublattice index while the $d$-wave terms are odd in the sublattice index. From Eq.~(\ref{eq:gap_basis}) we can see that, in the absence of the $p$-wave component, the order parameter is completely off-diagonal in sublattice space. 

To reduce the complexity of the problem we first neglect the $p$-wave component by setting $\delta=0$. This term has by far the smallest contribution and as an intra-sublattice term we do not expect it to interfere significantly with any potential odd-$\omega$ pairing originating from the inter-sublattice channels. This also allows us to make direct contact with previously-derived expressions for the Kerr effect in UPt$_3$ which also used $\delta=0$. 
In this case, the Hamiltonian in Eq.~(\ref{eq:hamiltonian}) breaks down into the following two decoupled sectors:\cite{wang_2017}
\begin{equation}
\begin{aligned}
\mathcal{H}&=\mathcal{H}_a+\mathcal{H}_b \\
&=\frac{1}{2}\sum_\textbf{k}\left[\Psi^\dagger_{a,\textbf{k}}\hat{\mathcal{H}}_a(\textbf{k})\Psi_{a,\textbf{k}}+\Psi^\dagger_{b,\textbf{k}}\hat{\mathcal{H}}_b(\textbf{k})\Psi_{b,\textbf{k}}\right]
\end{aligned}
\end{equation} 
where
\begin{equation}
\hat{\mathcal{H}}_a(\textbf{k})=\left(\begin{array}{cccc}
\xi_\textbf{k}+g_\textbf{k} & \epsilon_\textbf{k} & 0 & \Delta_{12}(\textbf{k}) \\
\epsilon^*_\textbf{k} & \xi_\textbf{k}-g_\textbf{k} & \Delta_{21}(\textbf{k}) & 0 \\
0 & \Delta^*_{21}(\textbf{k}) & -\xi_\textbf{k}-g_\textbf{k} & -\epsilon_\textbf{k} \\
\Delta^*_{12}(\textbf{k}) & 0 & -\epsilon^*_\textbf{k} & -\xi_\textbf{k}+g_\textbf{k} \\
\end{array} \right)
\label{eq:hama}
\end{equation}
\begin{equation}
\hat{\mathcal{H}}_b(\textbf{k})=\left(\begin{array}{cccc}
\xi_\textbf{k}-g_\textbf{k} & \epsilon_\textbf{k} & 0 & \Delta_{12}(\textbf{k}) \\
\epsilon^*_\textbf{k} & \xi_\textbf{k}+g_\textbf{k} & \Delta_{21}(\textbf{k}) & 0 \\
0 & \Delta^*_{21}(\textbf{k}) & -\xi_\textbf{k}+g_\textbf{k} & -\epsilon_\textbf{k} \\
\Delta^*_{12}(\textbf{k}) & 0 & -\epsilon^*_\textbf{k} & -\xi_\textbf{k}-g_\textbf{k} \\
\end{array} \right) 
\label{eq:hamb}
\end{equation}
and the gap functions take the form
\begin{equation}
\begin{aligned}
\Delta_{12}(\textbf{k})&=f_\textbf{k} + id_\textbf{k}, \\
\Delta_{21}(\textbf{k})&=f_\textbf{k} - id_\textbf{k}, \\
\end{aligned}
\label{eq:gap_ij}
\end{equation}
with 
\begin{equation}
\begin{aligned}
f_\textbf{k}&= \eta_1 f_{(x^2-y^2)z}(\textbf{k}) +\eta_2 f_{xyz}(\textbf{k}) , \\
d_\textbf{k}&= \eta_1 d_{yz}(\textbf{k}) +\eta_2 d_{xz}(\textbf{k}) , \\
\end{aligned}
\label{eq:fd}
\end{equation}
and all expressed with the bases defined by
\begin{equation}
\begin{aligned}
\Psi^\dagger_{a,\textbf{k}}&=\left(\begin{array}{cccc}
c^\dagger_{\textbf{k}1\uparrow} & c^\dagger_{\textbf{k}2\uparrow} & c_{-\textbf{k}1\downarrow} & c_{-\textbf{k}2\downarrow} 
\end{array} \right) \\
\Psi^\dagger_{b,\textbf{k}}&=\left(\begin{array}{cccc}
c^\dagger_{\textbf{k}1\downarrow} & c^\dagger_{\textbf{k}2\downarrow} & c_{-\textbf{k}1\uparrow} & c_{-\textbf{k}2\uparrow}
\end{array} \right). 
\end{aligned}
\end{equation}
Notice that the two 4$\times$4 Hamiltonian matrices are nearly identical, they are related by simply changing the sign of the spin-orbit coupling $g_\textbf{k}$. Therefore, without loss of generality, for the remainder of this article, we focus on just $\mathcal{H}_a$ and note that the results for $\mathcal{H}_b$ may be obtained by taking $\alpha\rightarrow-\alpha$.

\section{Pair Symmetry Classification}
\label{sec:pairsym}

We begin our discussion by defining both the normal Green's function, $G$, and the anomalous Green's function, $F$, in terms of the creation and annihilation operators in Eq.~(\ref{eq:hamiltonian}):
\begin{equation}
\begin{aligned}
G_{mm',\sigma\sigma'}(\textbf{k};\tau)&=-\langle T_\tau c_{\textbf{k}m\sigma}(\tau)c^\dagger_{\textbf{k}m'\sigma'}(0) \rangle, \\
F_{mm',\sigma\sigma'}(\textbf{k};\tau)&=-\langle T_\tau c_{\textbf{k}m\sigma}(\tau)c_{-\textbf{k}m'\sigma'}(0) \rangle, \\
\end{aligned}
\label{eq:g_tau}
\end{equation}
where $\tau$ is imaginary time and $T_\tau$ is the usual $\tau$-ordering operator for fermions. The pair symmetry of UPt$_3$ can be determined by studying the anomalous Green's function, $F$. 

Using the Hamiltonian matrix in Eq.~(\ref{eq:hama}) it is straightforward to find the Matsubara representation of the total Green's functions from 
\begin{equation}
\hat{\mathcal{G}}_a(\textbf{k};i\omega_n)=\left[i\omega_n - \hat{\mathcal{H}}_a(\textbf{k})\right]^{-1}
\label{eq:G_alpha}
\end{equation}
where $\omega_n=\pi(2n+1)/\beta$ is a Matsubara frequency for inverse temperature $\beta$. Here $\hat{\mathcal{G}}_a(\textbf{k};i\omega_n)$ is a 4$\times$4 matrix with components given by
\begin{equation}
\hat{\mathcal{G}}_a=\left(\begin{array}{cccc}
G_{11,\uparrow\uparrow} & G_{12,\uparrow\uparrow} & F_{11,\uparrow\downarrow} & F_{12,\uparrow\downarrow} \\
G_{21,\uparrow\uparrow} & G_{22,\uparrow\uparrow} & F_{21,\uparrow\downarrow} & F_{22,\uparrow\downarrow} \\
-F^*_{11,\downarrow\uparrow} & -F^*_{12,\downarrow\uparrow} & -G^*_{11,\downarrow\downarrow} & -G^*_{12,\downarrow\downarrow} \\
-F^*_{21,\downarrow\uparrow} & -F^*_{22,\downarrow\uparrow} & -G^*_{21,\downarrow\downarrow} & -G^*_{22,\downarrow\downarrow} \\
\end{array} \right),
\end{equation}
where we have suppressed the dependence on crystal momentum $\textbf{k}$ and Matsubara frequency $\omega_n$ for brevity. Notice that $\hat{\mathcal{G}}_a$ is comprised of four 2$\times$2 blocks in sublattice space, where we can identify the upper off-diagonal block as the anomalous Green's function: $\hat{F}_{a}\equiv \hat{F}_{\uparrow\downarrow}$.

To isolate $\hat{F}_a$, we start by noting that the Hamiltonian $\hat{\mathcal{H}_a}$ in Eq.~(\ref{eq:hama}) has the form:
\begin{equation}
\hat{\mathcal{H}_a}=\left(\begin{array}{cc}
\hat{h}_\textbf{k} & \hat{\Delta}(\textbf{k}) \\
\hat{\Delta}^\dagger(\textbf{k}) & -\hat{h}_\textbf{k}
\end{array} \right).
\label{eq:ham_blocks}
\end{equation}
Combining Eq.~(\ref{eq:ham_blocks}) with Eq.~(\ref{eq:G_alpha}) it can be shown that the anomalous Green's function associated with $\hat{\mathcal{H}_a}$ is given by
\begin{equation}
\begin{aligned}
\hat{F}_a(\textbf{k};i\omega_n)\!=\! \left[ (i\omega_n+\hat{h}_\textbf{k})\hat{\Delta}^{-1}(\textbf{k})(i\omega_n-\hat{h}_\textbf{k})-\hat{\Delta}^\dagger(\textbf{k})\right]^{-1}.
\end{aligned}
\label{eq:fa_matrix}
\end{equation}  

The remainder of this section is dedicated to understanding the symmetries of Eq.~(\ref{eq:fa_matrix}). 
However, before we study the general expressions for the full model appearing in Eq.~(\ref{eq:hama}) we turn our attention to three limiting cases: (i) no spin-orbit coupling or inter-sublattice hopping, $\alpha=t'=0$; (ii) no spin-orbit coupling but finite inter-sublattice hopping, $\alpha=0$, $t'\neq0$; and (iii) finite spin-orbit coupling but no inter-sublattice hopping, $\alpha\neq0$, $t'=0$.

\subsection{Case (i): $\alpha=t'=0$}
In the limit of no spin-orbit coupling and no inter-sublattice hopping, $\alpha=t'=0$, all terms proportional to $g_\textbf{k}$ or $\epsilon_\textbf{k}$ vanish in $\hat{\mathcal{H}_a}$. In this case Eq.~(\ref{eq:fa_matrix}) takes on a relatively simple form:
\begin{widetext}
\begin{equation}
\hat{F}_a(\textbf{k};i\omega_n)=A_{\textbf{k};i\omega_n}\left(\begin{array}{cc}
0 & \Delta_{12}(\textbf{k})\left[ (i\omega_n)^2-E_{21}(\textbf{k})^2 \right] \\
\Delta_{21}(\textbf{k})\left[ (i\omega_n)^2-E_{12}(\textbf{k})^2 \right] & 0
\end{array} \right)
\label{eq:F_(i)}
\end{equation}
where we have defined
\begin{equation}
\begin{aligned}
A_{\textbf{k};i\omega_n}^{-1}&=\left\{ (i\omega_n)^4-(i\omega_n)^2\left[E_{12}^2(\textbf{k})+E_{21}^2(\textbf{k})\right] +E_{12}^2(\textbf{k})E_{21}^2(\textbf{k})\right\}, \\
E_{ij}&=\sqrt{\xi_\textbf{k}^2+|\Delta_{ij}(\textbf{k})|^2}.
\end{aligned}
\label{eq:denominator_(i)}
\end{equation}
\end{widetext}

This trivial case demonstrates that without any mixing of the sublattice or spin degrees of freedom the anomalous Green's function has exactly the same symmetries as the underlying gap functions, $\Delta_{ij}(\textbf{k})$. It is thus clear that there are no odd-$\omega$ pair amplitudes in this case. This result is consistent with the previous results in multiband superconductors showing a required mixing of degrees of freedom for odd-$\omega$ pairing.\cite{black2013odd} 

\subsection{Case (ii): $\alpha= 0$, $t'\neq 0$}
In the limit of no spin-orbit coupling but finite inter-sublattice hopping, $\alpha= 0$, $t'\neq 0$, we neglect all terms proportional to $g_\textbf{k}$ appearing in Eq.~(\ref{eq:hama}); however, we must keep track of the inter-sublattice hopping terms, $\epsilon_\textbf{k}$, which are, in general, complex numbers. In this case, the anomalous Green's function becomes:
\begin{widetext}
\begin{equation}
\hat{F}_a(\textbf{k};i\omega_n)=B_{\textbf{k};i\omega_n}\left(\begin{array}{cc}
i\omega_n a_{-}(\textbf{k})+\xi_\textbf{k}a_{+}(\textbf{k}) & \Delta_{12}(\textbf{k})\left[ (i\omega_n)^2-E_{21}(\textbf{k})^2 \right]-\Delta_{21}(\textbf{k})\epsilon_\textbf{k}^2 \\
\Delta_{21}(\textbf{k})\left[ (i\omega_n)^2-E_{12}(\textbf{k})^2 \right]-\Delta_{12}(\textbf{k}){\epsilon^*_\textbf{k}}^2 & -i\omega_n a_{-}(\textbf{k})+\xi_\textbf{k}a_{+}(\textbf{k})
\end{array} \right)
\label{eq:F_(iii)}
\end{equation}
where we have defined
\begin{equation}
\begin{aligned}
B_{\textbf{k};i\omega_n}^{-1}&=\left\{ (i\omega_n)^4-(i\omega_n)^2\left[E_{12}^2(\textbf{k})+E_{21}^2(\textbf{k})+2|\epsilon_\textbf{k}|^2\right]+E_{12}^2(\textbf{k})E_{21}^2(\textbf{k})  +|\epsilon_\textbf{k}|^4-2\xi_\textbf{k}^2|\epsilon_\textbf{k}|^2 + 2\mathcal{R}e\left[\Delta_{12}(\textbf{k})\Delta^*_{21}(\textbf{k}){\epsilon^*_\textbf{k}}^2\right]\right\}, \\
a_{\pm}&=\epsilon_\textbf{k}\Delta_{21}(\textbf{k})\pm\epsilon^*_\textbf{k}\Delta_{12}(\textbf{k}), \\
E_{ij}&=\sqrt{\xi_\textbf{k}^2+|\Delta_{ij}(\textbf{k})|^2}.
\end{aligned}
\label{eq:denominator_(iii)}
\end{equation}
By inspecting Eqs.~(\ref{eq:F_(iii)}) and (\ref{eq:denominator_(iii)}) it is straightforward to show that the inter-sublattice hopping terms generate one odd-$\omega$ pairing channel, given by:
\begin{equation}
\begin{aligned}
\hat{F}_{a,odd}(\textbf{k};i\omega_n)&=B_{\textbf{k};i\omega_n}\left(\begin{array}{cc}
2\omega_n \left[d_{\textbf{k}}\Re e\left\{\epsilon_\textbf{k}\right\}-f_{\textbf{k}}\Im m\left\{\epsilon_\textbf{k}\right\}\right] & 0 \\
0 & -2\omega_n \left[d_{\textbf{k}}\Re e\left\{\epsilon_\textbf{k}\right\}-f_{\textbf{k}}\Im m\left\{\epsilon_\textbf{k}\right\}\right]
\end{array} \right). 
\end{aligned}
\label{eq:F_odd_(iii)}
\end{equation} 
\end{widetext}
Note that this odd-$\omega$ pair amplitude is proportional to the sublattice hybridization but that it is strictly diagonal in the sublattice index, i.e.~it is intra-sublattice pairing. Hence, starting from the initial state with only inter-sublattice pairing, Eq.~(\ref{eq:hama}), the addition of inter-sublattice hopping terms acts to mix the electronic degrees of freedom on the two sublattices allowing for the emergence of a finite intra-sublattice pair amplitude. From the form of Eq.~(\ref{eq:F_odd_(iii)}) we see that, generically, for any finite inter-sublattice hopping amplitude $\epsilon_\textbf{k}$ this odd-$\omega$ intra-sublattice pair amplitude will be nonzero. Furthermore, we see that, because $\Re e\left\{\epsilon_\textbf{k}\right\}=\Re e\left\{\epsilon_{-\textbf{k}}\right\}$ and $\Im m\left\{\epsilon_\textbf{k}\right\}=-\Im m\left\{\epsilon_{-\textbf{k}}\right\}$ the symmetry constraints imposed by Fermi-Dirac statistics are always satisfied, since the odd-$\omega$ state is spin-triplet and has even spatial parity.

Notice that, in the absence of the $d$-wave component, the two gaps are equal, $\Delta_{12}(\textbf{k})=\Delta_{21}(\textbf{k})$. In this case, for purely real values of $\epsilon_\textbf{k}$ the odd-$\omega$ term vanishes. This is consistent with the expectation that emergent odd-$\omega$ pairing in a multiband superconductor is proportional to the difference between the two gaps.\cite{black2013odd, komendova2015experimentally, komendova2017odd, triola_prb_2017} However, the addition of an imaginary component to the interband hybridization allows the generation of odd-$\omega$ pairing in multiband systems even when the two gaps are equal. This is distinct from all other forms of odd-$\omega$ pairing that have been previously discussed in multiband superconductors.\cite{black2013odd,komendova2015experimentally,komendova2017odd,triola_prb_2017}

\subsection{Case (iii): $\alpha\neq 0$, $t'=0$}
In the limit of finite spin-orbit coupling but no inter-sublattice hopping, $\alpha\neq 0$, $t'=0$, we neglect all terms proportional to $\epsilon_\textbf{k}$ appearing in Eq.~(\ref{eq:hama}); however, we must keep track of all terms proportional to $g_\textbf{k}$, which is a real-valued and odd function of momentum. In this case, the anomalous Green's function becomes:
\begin{widetext}
\begin{equation}
\hat{F}_a(\textbf{k};i\omega_n)=C_{\textbf{k};i\omega_n}\left(\begin{array}{cc}
0 & \Delta_{12}(\textbf{k})\left[ (i\omega_n+g_\textbf{k})^2-E_{21}(\textbf{k})^2 \right] \\
\Delta_{21}(\textbf{k})\left[ (i\omega_n-g_\textbf{k})^2-E_{12}(\textbf{k})^2 \right] & 0
\end{array} \right)
\label{eq:F_(ii)}
\end{equation}
where we have defined
\begin{equation}
\begin{aligned}
C_{\textbf{k};i\omega_n}^{-1}&=\left\{ (i\omega_n)^4-(i\omega_n)^2\left[E_{12}^2(\textbf{k})+E_{21}^2(\textbf{k})+2g_\textbf{k}^2\right] +E_{12}^2(\textbf{k})E_{21}^2(\textbf{k})+4g_\textbf{k}\omega_n\left[d_\textbf{k}f_\textbf{k}^*-f_\textbf{k}d_\textbf{k}^*\right]\right. \\
&+\left. g_\textbf{k}^2\left[g_\textbf{k}^2-E_{12}^2(\textbf{k})-E_{21}^2(\textbf{k})\right]\right\}, \\
E_{ij}&=\sqrt{\xi_\textbf{k}^2+|\Delta_{ij}(\textbf{k})|^2}.
\end{aligned}
\label{eq:denominator_(ii)}
\end{equation}

From Eq.~(\ref{eq:F_(ii)}) we see immediately that the addition of spin-orbit coupling induces an odd-$\omega$ pair amplitude proportional to $g_\textbf{k}$. Careful inspection of Eqs.~(\ref{eq:F_(ii)}) and (\ref{eq:denominator_(ii)}) reveals that the addition of spin-orbit coupling in fact gives rise to two different terms contributing to odd-$\omega$ pairing, one from the numerator:
\begin{equation}
\hat{F}^{(1)}_{a,odd}(\textbf{k};i\omega_n)=C_{\textbf{k};i\omega_n}^{(1)}\left(\begin{array}{cc}
0 & 2i\omega_n g_\textbf{k}\left(f_\textbf{k}+id_\textbf{k}\right) \\
-2i\omega_n g_\textbf{k}\left(f_\textbf{k}-id_\textbf{k}\right) & 0
\end{array} \right),
\label{eq:F_odd_num_(ii)}
\end{equation} 
and one coming from an odd-$\omega$ term in the denominator:
\begin{equation}
\hat{F}^{(2)}_{a,odd}(\textbf{k};i\omega_n)=-4g_\textbf{k}\omega_n\left[d_\textbf{k}f_\textbf{k}^*-f_\textbf{k}d_\textbf{k}^*\right]\hat{F}_{a,even}(\textbf{k};i\omega_n),
\label{eq:F_odd_den_(ii)}
\end{equation}
where we have defined:
\begin{equation}
\hat{F}_{a,even}(\textbf{k};i\omega_n)=C_{\textbf{k};i\omega_n}^{(2)}\left(\begin{array}{cc}
0 & \Delta_{12}(\textbf{k})\left[ (i\omega_n)^2+g_\textbf{k}^2-E_{21}(\textbf{k})^2 \right] \\
\Delta_{21}(\textbf{k})\left[ (i\omega_n)^2+g_\textbf{k}^2-E_{12}(\textbf{k})^2 \right] & 0
\end{array} \right)
\label{eq:F_even_(ii)}
\end{equation}
\end{widetext}
where $C_{\textbf{k};i\omega_n}^{(1)}$ and $C_{\textbf{k};i\omega_n}^{(2)}$ are both strictly even functions of the frequency $\omega_n$ and can thus be ignored in the symmetry analysis. Notice that the spin-orbit coupling cannot modify the pairing in the sublattice index and thus we find that all pairing remains off-diagonal in the sublattice index, just as in case (i). 

Recalling that $g_{\textbf{k}}=-g_{-\textbf{k}}$ we can check that the constraints imposed by Fermi-Dirac statistics are satisfied in Eqs.~(\ref{eq:F_odd_num_(ii)}) and (\ref{eq:F_odd_den_(ii)}). Inspection of Eq.~(\ref{eq:F_odd_num_(ii)}) shows that the $f$-wave term is converted from: sublattice-even to sublattice-odd; parity-odd to parity-even; and from even-$\omega$ to odd-$\omega$. Naively, this appears to violate the statistics; however, we must recall that these results only apply for the $a$ sector. To obtain the results for the $b$ sector we take $g_\textbf{k}\rightarrow-g_\textbf{k}$ and $\uparrow\leftrightarrow\downarrow$. Therefore, we see that the odd-$\omega$ terms in Eqs.~(\ref{eq:F_odd_num_(ii)}) and (\ref{eq:F_odd_den_(ii)}) are actually spin-singlet and hence Fermi-Dirac statistics are satisfied for both the $f_{\bf k}$ and $d_{\bf k}$ components.  

The first type of odd-$\omega$ pair amplitude, $\hat{F}^{(1)}_{a,odd}$, has its origin in the numerators appearing in Eq.~(\ref{eq:F_(ii)}). The emergence of this pair amplitude can be understood by noting that the spin-orbit coupling acts as a momentum-dependent exchange field modifying the spin-symmetry of the superconducting correlations and thereby generating terms in the anomalous Green's function that are odd in frequency. This is similar in spirit to previous analyses of heterostructures incorporating superconductors and materials with different kinds of spin-orbit coupling.\cite{triola2016prl} 

The second type of odd-$\omega$ pair amplitude, $\hat{F}^{(2)}_{a,odd}$, has its origin in the denominator appearing in Eq.~(\ref{eq:denominator_(ii)}) and is proportional to $g_\textbf{k}\omega_n\left[d_\textbf{k}f_\textbf{k}^*-f_\textbf{k}d_\textbf{k}^*\right]$. While the contribution from the numerator, $\hat{F}^{(1)}_{a,odd}$, is, in general, nonzero for any finite spin-orbit coupling, the term coming from the denominator, $\hat{F}^{(2)}_{a,odd}$, requires both a finite spin-orbit coupling and a gap such that, $d_\textbf{k}f_\textbf{k}^*-f_\textbf{k}d_\textbf{k}^*\neq0$. Considering the form of $f_\textbf{k}$ and $d_\textbf{k}$ given in Eq.~(\ref{eq:fd}), we conclude that this term is only nonzero in the non-unitary $B$ phase of UPt$_3$, where both $\eta_1$ and $\eta_2$ take on finite values.

\subsection{General case: $\alpha,t'\neq 0$}

Finally, we turn our attention to the most general case, both finite spin-orbit coupling and inter-sublattice hopping, $\alpha,t'\neq 0$. In this case the anomalous Green's function is given by:
\begin{widetext}
\begin{equation}
\hat{F}_a(\textbf{k};i\omega_n)=D_{\textbf{k};i\omega_n}\left(\begin{array}{cc}
i\omega_n a_{-}(\textbf{k})+(\xi_\textbf{k}-g_\textbf{k})a_{+}(\textbf{k}) & \Delta_{12}(\textbf{k})\left[ (i\omega_n+g_\textbf{k})^2-E_{21}(\textbf{k})^2 \right]-\Delta_{21}(\textbf{k})\epsilon_\textbf{k}^2 \\
\Delta_{21}(\textbf{k})\left[ (i\omega_n-g_\textbf{k})^2-E_{12}(\textbf{k})^2 \right]-\Delta_{12}(\textbf{k}){\epsilon^*_\textbf{k}}^2 & -i\omega_n a_{-}(\textbf{k})+(\xi_\textbf{k}+g_\textbf{k})a_{+}(\textbf{k})
\end{array} \right)
\label{eq:F}
\end{equation}
where we have defined
\begin{equation}
\begin{aligned}
D_{\textbf{k};i\omega_n}^{-1}&=\left\{ (i\omega_n)^4-2(i\omega_n)^2\left[\xi_\textbf{k}^2+|\epsilon_\textbf{k}|^2+g_\textbf{k}^2+|f_\textbf{k}|^2+|d_\textbf{k}|^2\right] +4g_\textbf{k}\omega_n\left[d_\textbf{k}f_\textbf{k}^*-f_\textbf{k}d_\textbf{k}^*\right]\right. \\
&+\left. \xi_\textbf{k}^4+|\epsilon_\textbf{k}|^4+g_\textbf{k}^4-2g_\textbf{k}^2\xi_\textbf{k}^2+2(\xi_\textbf{k}^2-g_\textbf{k}^2)\left[|f_\textbf{k}|^2+|d_\textbf{k}|^2-|\epsilon_\textbf{k}|^2\right]+|\Delta_{12}(\textbf{k})|^2|\Delta_{21}(\textbf{k})|^2 \right. \\
&+\left. 2\Re e\left[\Delta_{12}(\textbf{k})\Delta^*_{21}(\textbf{k}){\epsilon^*_\textbf{k}}^2\right]\right\}, \\
a_{\pm}&=\epsilon_\textbf{k}\Delta_{21}(\textbf{k})\pm\epsilon^*_\textbf{k}\Delta_{12}(\textbf{k}), \\
E_{ij}&=\sqrt{\xi_\textbf{k}^2+|\Delta_{ij}(\textbf{k})|^2}.
\end{aligned}
\label{eq:denominator}
\end{equation}
By inspecting Eqs.~(\ref{eq:F}) and (\ref{eq:denominator}), once again, we see the emergence of two distinct kinds of odd-$\omega$ pairing amplitudes, one from the numerator:
\begin{equation}
\hat{F}^{(1)}_{a,odd}(\textbf{k};i\omega_n)=D_{\textbf{k};i\omega_n}^{(1)}\left(\begin{array}{cc}
2\omega_n \left[d_{\textbf{k}}\Re e\left\{\epsilon_\textbf{k}\right\}-f_{\textbf{k}}\Im m\left\{\epsilon_\textbf{k}\right\}\right] & 2i\omega_n g_\textbf{k}\left(f_\textbf{k}+id_\textbf{k}\right) \\
-2i\omega_n g_\textbf{k}\left(f_\textbf{k}-id_\textbf{k}\right) & -2\omega_n \left[d_{\textbf{k}}\Re e\left\{\epsilon_\textbf{k}\right\}-f_{\textbf{k}}\Im m\left\{\epsilon_\textbf{k}\right\}\right]
\end{array} \right), 
\label{eq:F_odd_num}
\end{equation} 
and one from the denominator:
\begin{equation}
\hat{F}^{(2)}_{a,odd}(\textbf{k};i\omega_n)=-4 g_\textbf{k}\omega_n\left[d_\textbf{k}f_\textbf{k}^*-f_\textbf{k}d_\textbf{k}^*\right]\hat{F}_{a,even}(\textbf{k};i\omega_n),
\label{eq:F_odd_den}
\end{equation}
where we define:
\begin{equation}
\hat{F}_{a,even}(\textbf{k};i\omega_n)=D_{\textbf{k};i\omega_n}^{(2)}\left(\begin{array}{cc}
(\xi_\textbf{k}-g_\textbf{k})\left[\epsilon_\textbf{k}\Delta_{21}(\textbf{k})+\epsilon^*_\textbf{k}\Delta_{12}(\textbf{k})\right] & \Delta_{12}(\textbf{k})\left[ (i\omega_n)^2+g_\textbf{k}^2-E_{21}(\textbf{k})^2 \right]-\Delta_{21}(\textbf{k})\epsilon_\textbf{k}^2 \\
\Delta_{21}(\textbf{k})\left[ (i\omega_n)^2+g_\textbf{k}^2-E_{12}(\textbf{k})^2 \right]-\Delta_{12}(\textbf{k}){\epsilon^*_\textbf{k}}^2 & (\xi_\textbf{k}+g_\textbf{k})\left[\epsilon_\textbf{k}\Delta_{21}(\textbf{k})+\epsilon^*_\textbf{k}\Delta_{12}(\textbf{k})\right]
\end{array} \right)
\label{eq:F_even}
\end{equation}
\end{widetext}
where the functions $D_{\textbf{k};i\omega_n}^{(1)}$ and $D_{\textbf{k};i\omega_n}^{(2)}$ are strictly even functions of frequency $\omega_n$. Overall, the relative sizes of these odd-$\omega$ terms depend on the parameters: the spin-orbit coupling term, $g_\textbf{k}$, the band hybridization, $\epsilon_\textbf{k}$, and gap symmetry parameterized by $\eta_1$ and $\eta_2$. 
This general result is a combination of cases (ii) and (iii) with additional terms appearing in $\hat{F}_{a,even}$ due to $g_{\bf k}$ and $\epsilon_{\bf k}$ being non-zero simultaneously.

From Eqs.~(\ref{eq:F_odd_num}) and (\ref{eq:F_odd_den}) we can deduce the general criteria for obtaining finite odd-$\omega$ pair amplitudes in UPt$_3$. In order to obtain finite odd-$\omega$ \textit{inter-sublattice} pair amplitudes we only need $g_\textbf{k}\neq 0$. Additional terms are present when $d_\textbf{k}f_\textbf{k}^*-f_\textbf{k}d_\textbf{k}^*\neq 0$; however, these latter terms do not lead to distinct channels in the sublattice index or in spatial parity. To obtain finite odd-$\omega$ \textit{intra-sublattice} pair amplitudes only a finite inter-sublattice hopping term is necessary, $\epsilon_\textbf{k}\neq 0$. This odd-$\omega$ intra-sublattice channel is predominantly $f$-wave when $\epsilon_\textbf{k}$ is purely imaginary and $d$-wave when $\epsilon_\textbf{k}$ is purely real. It also has additional contributions when both $d_\textbf{k}f_\textbf{k}^*-f_\textbf{k}d_\textbf{k}^*$ and $g_{\bf k}$ are non-zero

Having derived the general criteria for odd-$\omega$ pairing we finally focus on the specific Fermi surfaces.
First we recall that the Fermi surface of UPt$_3$ around the $\Gamma$ point is describable with parameters: $\left(t,t_z,t',\alpha,\mu\right)=\left(1,4,1,0,16\right)$. Therefore, we conclude that near the $\Gamma$ point in UPt$_3$ we expect to find odd-$\omega$ pairing proportional to $\epsilon_\textbf{k}$, and only in the intra-sublattice channel since $g_{\bf k} = 0$. Furthermore, these amplitudes are present in all three phases, $A$, $B$, and $C$, since they appear very generally for any finite values of $\eta_1$,$\eta_2$ in Eq (\ref{eq:fd}).
Next, recall the set of parameters describing UPt$_3$ around the $A$ point: $\left(t,t_z,t',\alpha,\mu\right)=\left(1,-4,1,2,12\right)$. From this parameter set we see that near the $A$ point in UPt$_3$ both the odd-$\omega$ intra-sublattice channel and odd-$\omega$ inter-sublattice channel are finite in all three phases. Furthermore, while both the $A$ phase and $C$ phase are unitary according to our model, the $B$ phase is not. This results in the $B$ phase receiving additional contributions of the form given in Eq.~(\ref{eq:F_odd_den}) in both the odd-$\omega$ intra-sublattice and odd-$\omega$ inter-sublattice channels. These additional contributions, originating from a term in the denominator being odd in frequency, are absent in the $A$ and $C$ phases.   

It should be noted that, while the crystal symmetry of UPt$_3$ was originally recorded to be the close-packed hexagonal symmetry of space group $P6_3/mmc$\cite{heal_1955,walko_2001_prb} there has been an indication in measurements combining x-ray diffraction and transition electron microscopy that the actual symmetry may be that of the trigonal space group $P\overline{3}m1$.\cite{walko_2001_prb} In this case, the lattice will be distorted from the picture appearing in Fig.~\ref{fig:lattice} leading to a layer dimerization that will introduce an asymmetry in the inter-sublattice hybridization along the $z$-axis. In the presence of this asymmetry the inter-sublattice hopping term, $\epsilon_{\textbf{k}}$, becomes\cite{yanase_2017_prb}
\begin{equation}
\begin{aligned}
\tilde{\epsilon}_{\textbf{k}}&=t'\left[(1+d)e^{i\tfrac{k_z}{2}}+(1-d)e^{-i\tfrac{k_z}{2}} \right]\sum_ie^{i\textbf{k}_{||}\cdot\textbf{r}_i} \\
&=2t'\left[\cos\tfrac{k_z}{2} + id\sin\tfrac{k_z}{2}\right]  \sum_ie^{i\textbf{k}_{||}\cdot\textbf{r}_i}
\end{aligned}
\label{eq:dimerization}
\end{equation}
where $d$ parameterizes the magnitude of the layer dimerization. From this expression we note that, just as with $\epsilon_{\textbf{k}}$, $\Re e\{\tilde{\epsilon}_\textbf{k}\}$ is strictly even in $\textbf{k}$, while $\Im m\{\tilde{\epsilon}_\textbf{k}\}$ is strictly odd in $\textbf{k}$. Hence, while the precise momentum dependence of the odd-$\omega$ pairing will be affected by the crystal distortion described by Eq.~(\ref{eq:dimerization}), the qualitative features discussed above for the odd-$\omega$ pair amplitudes emerging in the absence of the crystal distortion will be preserved even in the presence of such a crystal distortion.

\section{Odd-frequency Pairing and the Kerr Effect}
\label{sec:kerr}
Now that we have shown how odd-$\omega$ superconductivity is ubiquitous in UPt$_3$, we turn to investigating its relationship with the Kerr effect, which has been measured within the $B$ phase. In general, the frequency-dependent rotation angle for the polarization of reflected light, known as the Kerr angle, is given by:
\begin{equation}
\theta(\omega)=\frac{4\pi}{\omega}\Im m\left[\frac{\sigma_H(\omega)}{n(n^2-1)} \right]
\label{eq:kerr}
\end{equation}
where $\sigma_H(\omega)$ is the anomalous Hall conductivity and $n$ is the index of refraction for the material. Motivated by observations of a finite Kerr angle in UPt$_3$,\cite{schemm_2014} a recent work\cite{wang_2017} computed the value of Eq.~(\ref{eq:kerr}) employing the same Hamiltonian as used here, Eq.~(\ref{eq:hamiltonian}). It was shown that $\theta(\omega)$ is given by a sum over the Brillouin zone of a quantity proportional to $[f_\textbf{k}d_\textbf{k}^*-f_\textbf{k}^*d_\textbf{k}]$, where $f_\textbf{k},d_\textbf{k}$ are those given by Eq.~(\ref{eq:fd}). Furthermore, it was determined that $\theta(\omega)=0$ if the inter-sublattice hopping function $\epsilon_\textbf{k}$ is real. Therefore, the criteria for observing a finite Kerr rotation angle in UPt$_3$ are given by: (1) $[f_\textbf{k}d_\textbf{k}^*-f_\textbf{k}^*d_\textbf{k}]\neq 0$, and (2) $\Im m\left\{\epsilon_\textbf{k}\right\}\neq 0$. Additionally, it was found that a second contribution to the Kerr angle arises when $g_\textbf{k}$ is finite. Comparing these criteria to the results of the previous sections of this work, we see that, while they differ from our general criteria for the emergence of odd-$\omega$ pairing amplitudes in UPt$_3$, there are still strong similarities.

In particular, notice that in order to observe the Kerr effect the system must possess finite $\epsilon_\textbf{k}$ and finite contributions from both $f_\textbf{k}$ and $d_\textbf{k}$. From this alone we can conclude that the system must then also possess odd-$\omega$ intra-sublattice pairing of the form appearing in Eq.~(\ref{eq:F_odd_(iii)}). Additionally, we note that when $g_\textbf{k}\neq 0$, the Kerr angle picks up an additional contribution, comparable in magnitude to the contribution independent of spin-orbit coupling.\cite{wang_2017} Directly correlated with this additional term is the emergence of a finite odd-$\omega$ inter-sublattice pairing amplitude with finite $g_{\bf k}$ given in Eqs.~\eqref{eq:F_odd_num_(ii)} and \eqref{eq:F_odd_den_(ii)}. Also, since this second term in $\sigma_H(\omega)$ is proportional to both $f_\textbf{k}d_\textbf{k}^*-f_\textbf{k}^*d_\textbf{k}$ as well as $g_\textbf{k}$, we see from Eq.~(\ref{eq:F_odd_den}) that novel odd-$\omega$ contributions to both the intra-sublattice and inter-sublattice pairing channels emerge, arising from the denominator of the Green's function. 

Therefore, we conclude that the observation of the Kerr effect directly implies the presence of odd-$\omega$ intra-sublattice pair amplitudes in UPt$_3$. While this observation alone cannot tell us whether or not odd-$\omega$ inter-sublattice pair amplitudes also exist in UPt$_3$, the same expressions determining the size of the odd-$\omega$ inter-sublattice pair amplitudes also provide a contribution to the size of the Kerr rotation angle.

\section{Conclusions}
\label{sec:con}

In this work we studied the emergence of odd-$\omega$ pairing in the heavy-fermion superconductor UPt$_3$. Using a tight-binding model describing the electrons associated with the U ions capturing the nonsymmorphic crystal structure, and assuming an order parameter belonging to $E_{2u}$, we characterized the emergence of odd-$\omega$ pair amplitudes and their dependence on the underlying parameters. We found that, in the presence of inter-sublattice hopping, odd-$\omega$ intra-sublattice pairing is present in the system in all three superconducting phases of UPt$_3$, $A$, $B$, and $C$. Similarly, in the presence of spin-orbit coupling we found that odd-$\omega$ inter-sublattice pairing will be present in all three superconducting phases. Furthermore, we found that in the $B$ phase additional odd-$\omega$ contributions emerge. Considering model parameters which faithfully portray the topology of the Fermi surface near the $\Gamma$ point and $A$ point, we found that near $\Gamma$ the model predicts finite odd-$\omega$ intra-sublattice pairing, while at $A$ both odd-$\omega$ intra-sublattice and odd-$\omega$ inter-sublattice pairing are finite. Additionally, we compared our criteria for the realization of odd-$\omega$ pairing in UPt$_3$ to recent calculations of the size of the Kerr effect using the same model\cite{wang_2017} and found very strong similarities. Notably, we showed that when the Kerr rotation angle is finite, odd-$\omega$ pair amplitudes are always present. Since the Kerr effect has been observed in UPt$_3$,\cite{schemm_2014} this strongly suggests the presence of odd-$\omega$ pairing in this heavy-fermion superconductor.

\acknowledgments 
We thank A. V. Balatsky, A. Bouhon, R. M. Geilhufe, Y. Kedem, and L. Komendova for useful discussions. This work was supported by the Swedish Research Council (Vetenskapsr\aa det) Grant No. 621-2014-3721, and the Knut and Alice Wallenberg Foundation through the Wallenberg Academy Fellows program.
    
\bibliographystyle{apsrevmy}
\bibliography{Odd_Frequency}

\begin{thebibliography}{74}
\expandafter\ifx\csname natexlab\endcsname\relax\def\natexlab#1{#1}\fi
\expandafter\ifx\csname bibnamefont\endcsname\relax
  \def\bibnamefont#1{#1}\fi
\expandafter\ifx\csname bibfnamefont\endcsname\relax
  \def\bibfnamefont#1{#1}\fi
\expandafter\ifx\csname citenamefont\endcsname\relax
  \def\citenamefont#1{#1}\fi
\expandafter\ifx\csname url\endcsname\relax
  \def\url#1{\texttt{#1}}\fi
\expandafter\ifx\csname urlprefix\endcsname\relax\def\urlprefix{URL }\fi
\providecommand{\bibinfo}[2]{#2}
\providecommand{\eprint}[2][]{\url{#2}}

\bibitem[{\citenamefont{Stewart et~al.}(1984)\citenamefont{Stewart, Fisk,
  Willis, and Smith}}]{stewart_prl_1984}
\bibinfo{author}{\bibfnamefont{G.~R.} \bibnamefont{Stewart}},
  \bibinfo{author}{\bibfnamefont{Z.}~\bibnamefont{Fisk}},
  \bibinfo{author}{\bibfnamefont{J.~O.} \bibnamefont{Willis}},
  \bibnamefont{and} \bibinfo{author}{\bibfnamefont{J.~L.} \bibnamefont{Smith}},
  \bibinfo{journal}{Phys. Rev. Lett.} \textbf{\bibinfo{volume}{52}},
  \bibinfo{pages}{679} (\bibinfo{year}{1984}).

\bibitem[{\citenamefont{Sauls}(1994)}]{sauls1994}
\bibinfo{author}{\bibfnamefont{J.}~\bibnamefont{Sauls}}, \bibinfo{journal}{Adv.
  Phys.} \textbf{\bibinfo{volume}{43}}, \bibinfo{pages}{113}
  (\bibinfo{year}{1994}).

\bibitem[{\citenamefont{Adenwalla et~al.}(1990)\citenamefont{Adenwalla, Lin,
  Ran, Zhao, Ketterson, Sauls, Taillefer, Hinks, Levy, and
  Sarma}}]{adenwalla1990}
\bibinfo{author}{\bibfnamefont{S.}~\bibnamefont{Adenwalla}},
  \bibinfo{author}{\bibfnamefont{S.~W.} \bibnamefont{Lin}},
  \bibinfo{author}{\bibfnamefont{Q.~Z.} \bibnamefont{Ran}},
  \bibinfo{author}{\bibfnamefont{Z.}~\bibnamefont{Zhao}},
  \bibinfo{author}{\bibfnamefont{J.~B.} \bibnamefont{Ketterson}},
  \bibinfo{author}{\bibfnamefont{J.~A.} \bibnamefont{Sauls}},
  \bibinfo{author}{\bibfnamefont{L.}~\bibnamefont{Taillefer}},
  \bibinfo{author}{\bibfnamefont{D.~G.} \bibnamefont{Hinks}},
  \bibinfo{author}{\bibfnamefont{M.}~\bibnamefont{Levy}}, \bibnamefont{and}
  \bibinfo{author}{\bibfnamefont{B.~K.} \bibnamefont{Sarma}},
  \bibinfo{journal}{Phys. Rev. Lett.} \textbf{\bibinfo{volume}{65}},
  \bibinfo{pages}{2298} (\bibinfo{year}{1990}).

\bibitem[{\citenamefont{Tou et~al.}(1996)\citenamefont{Tou, Kitaoka, Asayama,
  Kimura, \ifmmode~\bar{O}\else \={O}\fi{}nuki, Yamamoto, and
  Maezawa}}]{tou_prl_1996}
\bibinfo{author}{\bibfnamefont{H.}~\bibnamefont{Tou}},
  \bibinfo{author}{\bibfnamefont{Y.}~\bibnamefont{Kitaoka}},
  \bibinfo{author}{\bibfnamefont{K.}~\bibnamefont{Asayama}},
  \bibinfo{author}{\bibfnamefont{N.}~\bibnamefont{Kimura}},
  \bibinfo{author}{\bibfnamefont{Y.}~\bibnamefont{\ifmmode~\bar{O}\else
  \={O}\fi{}nuki}}, \bibinfo{author}{\bibfnamefont{E.}~\bibnamefont{Yamamoto}},
  \bibnamefont{and} \bibinfo{author}{\bibfnamefont{K.}~\bibnamefont{Maezawa}},
  \bibinfo{journal}{Phys. Rev. Lett.} \textbf{\bibinfo{volume}{77}},
  \bibinfo{pages}{1374} (\bibinfo{year}{1996}).

\bibitem[{\citenamefont{Strand et~al.}(2010)\citenamefont{Strand, Bahr,
  Van~Harlingen, Davis, Gannon, and Halperin}}]{strand_science_2010}
\bibinfo{author}{\bibfnamefont{J.~D.} \bibnamefont{Strand}},
  \bibinfo{author}{\bibfnamefont{D.~J.} \bibnamefont{Bahr}},
  \bibinfo{author}{\bibfnamefont{D.~J.} \bibnamefont{Van~Harlingen}},
  \bibinfo{author}{\bibfnamefont{J.~P.} \bibnamefont{Davis}},
  \bibinfo{author}{\bibfnamefont{W.~J.} \bibnamefont{Gannon}},
  \bibnamefont{and} \bibinfo{author}{\bibfnamefont{W.~P.}
  \bibnamefont{Halperin}}, \bibinfo{journal}{Science}
  \textbf{\bibinfo{volume}{328}}, \bibinfo{pages}{1368} (\bibinfo{year}{2010}).

\bibitem[{\citenamefont{Strand et~al.}(2009)\citenamefont{Strand,
  Van~Harlingen, Kycia, and Halperin}}]{strand_prl_2009}
\bibinfo{author}{\bibfnamefont{J.~D.} \bibnamefont{Strand}},
  \bibinfo{author}{\bibfnamefont{D.~J.} \bibnamefont{Van~Harlingen}},
  \bibinfo{author}{\bibfnamefont{J.~B.} \bibnamefont{Kycia}}, \bibnamefont{and}
  \bibinfo{author}{\bibfnamefont{W.~P.} \bibnamefont{Halperin}},
  \bibinfo{journal}{Phys. Rev. Lett.} \textbf{\bibinfo{volume}{103}},
  \bibinfo{pages}{197002} (\bibinfo{year}{2009}).

\bibitem[{\citenamefont{Schemm et~al.}(2014)\citenamefont{Schemm, Gannon,
  Wishne, Halperin, and Kapitulnik}}]{schemm_2014}
\bibinfo{author}{\bibfnamefont{E.~R.} \bibnamefont{Schemm}},
  \bibinfo{author}{\bibfnamefont{W.~J.} \bibnamefont{Gannon}},
  \bibinfo{author}{\bibfnamefont{C.~M.} \bibnamefont{Wishne}},
  \bibinfo{author}{\bibfnamefont{W.~P.} \bibnamefont{Halperin}},
  \bibnamefont{and}
  \bibinfo{author}{\bibfnamefont{A.}~\bibnamefont{Kapitulnik}},
  \bibinfo{journal}{Science} \textbf{\bibinfo{volume}{345}},
  \bibinfo{pages}{190} (\bibinfo{year}{2014}).

\bibitem[{\citenamefont{Nomoto and Ikeda}(2016)}]{nomoto_prl_2016}
\bibinfo{author}{\bibfnamefont{T.}~\bibnamefont{Nomoto}} \bibnamefont{and}
  \bibinfo{author}{\bibfnamefont{H.}~\bibnamefont{Ikeda}},
  \bibinfo{journal}{Phys. Rev. Lett.} \textbf{\bibinfo{volume}{117}},
  \bibinfo{pages}{217002} (\bibinfo{year}{2016}).

\bibitem[{\citenamefont{Yanase}(2016)}]{yanase_prb_2016}
\bibinfo{author}{\bibfnamefont{Y.}~\bibnamefont{Yanase}},
  \bibinfo{journal}{Phys. Rev. B} \textbf{\bibinfo{volume}{94}},
  \bibinfo{pages}{174502} (\bibinfo{year}{2016}).

\bibitem[{\citenamefont{Yanase and Shiozaki}(2017)}]{yanase_2017_prb}
\bibinfo{author}{\bibfnamefont{Y.}~\bibnamefont{Yanase}} \bibnamefont{and}
  \bibinfo{author}{\bibfnamefont{K.}~\bibnamefont{Shiozaki}},
  \bibinfo{journal}{Phys. Rev. B} \textbf{\bibinfo{volume}{95}},
  \bibinfo{pages}{224514} (\bibinfo{year}{2017}).

\bibitem[{\citenamefont{Wang et~al.}(2017)\citenamefont{Wang, Berlinsky,
  Zwicknagl, and Kallin}}]{wang_2017}
\bibinfo{author}{\bibfnamefont{Z.}~\bibnamefont{Wang}},
  \bibinfo{author}{\bibfnamefont{J.}~\bibnamefont{Berlinsky}},
  \bibinfo{author}{\bibfnamefont{G.}~\bibnamefont{Zwicknagl}},
  \bibnamefont{and} \bibinfo{author}{\bibfnamefont{C.}~\bibnamefont{Kallin}},
  \bibinfo{journal}{Phys. Rev. B} \textbf{\bibinfo{volume}{96}},
  \bibinfo{pages}{174511} (\bibinfo{year}{2017}).

\bibitem[{\citenamefont{Berezinskii}(1974)}]{Berezinskii1974}
\bibinfo{author}{\bibfnamefont{V.~L.} \bibnamefont{Berezinskii}},
  \bibinfo{journal}{Pis'ma Zh. Eksp. Teor. Fiz.} \textbf{\bibinfo{volume}{20}},
  \bibinfo{pages}{628} (\bibinfo{year}{1974}).

\bibitem[{\citenamefont{Kirkpatrick and Belitz}(1991)}]{kirkpatrick_1991_prl}
\bibinfo{author}{\bibfnamefont{T.~R.} \bibnamefont{Kirkpatrick}}
  \bibnamefont{and} \bibinfo{author}{\bibfnamefont{D.}~\bibnamefont{Belitz}},
  \bibinfo{journal}{Phys. Rev. Lett.} \textbf{\bibinfo{volume}{66}},
  \bibinfo{pages}{1533} (\bibinfo{year}{1991}).

\bibitem[{\citenamefont{Belitz and Kirkpatrick}(1992)}]{belitz_1992_prb}
\bibinfo{author}{\bibfnamefont{D.}~\bibnamefont{Belitz}} \bibnamefont{and}
  \bibinfo{author}{\bibfnamefont{T.~R.} \bibnamefont{Kirkpatrick}},
  \bibinfo{journal}{Phys. Rev. B} \textbf{\bibinfo{volume}{46}},
  \bibinfo{pages}{8393} (\bibinfo{year}{1992}).

\bibitem[{\citenamefont{Balatsky and Abrahams}(1992)}]{BalatskyPRB1992}
\bibinfo{author}{\bibfnamefont{A.}~\bibnamefont{Balatsky}} \bibnamefont{and}
  \bibinfo{author}{\bibfnamefont{E.}~\bibnamefont{Abrahams}},
  \bibinfo{journal}{Phys. Rev. B} \textbf{\bibinfo{volume}{45}},
  \bibinfo{pages}{13125} (\bibinfo{year}{1992}).

\bibitem[{\citenamefont{Coleman et~al.}(1993)\citenamefont{Coleman, Miranda,
  and Tsvelik}}]{coleman_1993_prl}
\bibinfo{author}{\bibfnamefont{P.}~\bibnamefont{Coleman}},
  \bibinfo{author}{\bibfnamefont{E.}~\bibnamefont{Miranda}}, \bibnamefont{and}
  \bibinfo{author}{\bibfnamefont{A.}~\bibnamefont{Tsvelik}},
  \bibinfo{journal}{Phys. Rev. Lett.} \textbf{\bibinfo{volume}{70}},
  \bibinfo{pages}{2960} (\bibinfo{year}{1993}).

\bibitem[{\citenamefont{Coleman et~al.}(1994)\citenamefont{Coleman, Miranda,
  and Tsvelik}}]{coleman_1994_prb}
\bibinfo{author}{\bibfnamefont{P.}~\bibnamefont{Coleman}},
  \bibinfo{author}{\bibfnamefont{E.}~\bibnamefont{Miranda}}, \bibnamefont{and}
  \bibinfo{author}{\bibfnamefont{A.}~\bibnamefont{Tsvelik}},
  \bibinfo{journal}{Phys. Rev. B} \textbf{\bibinfo{volume}{49}},
  \bibinfo{pages}{8955} (\bibinfo{year}{1994}).

\bibitem[{\citenamefont{Coleman et~al.}(1995)\citenamefont{Coleman, Miranda,
  and Tsvelik}}]{coleman_1995_prl}
\bibinfo{author}{\bibfnamefont{P.}~\bibnamefont{Coleman}},
  \bibinfo{author}{\bibfnamefont{E.}~\bibnamefont{Miranda}}, \bibnamefont{and}
  \bibinfo{author}{\bibfnamefont{A.}~\bibnamefont{Tsvelik}},
  \bibinfo{journal}{Phys. Rev. Lett.} \textbf{\bibinfo{volume}{74}},
  \bibinfo{pages}{1653} (\bibinfo{year}{1995}).

\bibitem[{\citenamefont{Heid}(1995)}]{heid1995thermodynamic}
\bibinfo{author}{\bibfnamefont{R.}~\bibnamefont{Heid}}, \bibinfo{journal}{Z.
  Phys. B} \textbf{\bibinfo{volume}{99}}, \bibinfo{pages}{15}
  (\bibinfo{year}{1995}).

\bibitem[{\citenamefont{Belitz and Kirkpatrick}(1999)}]{belitz_1999_prb}
\bibinfo{author}{\bibfnamefont{D.}~\bibnamefont{Belitz}} \bibnamefont{and}
  \bibinfo{author}{\bibfnamefont{T.~R.} \bibnamefont{Kirkpatrick}},
  \bibinfo{journal}{Phys. Rev. B} \textbf{\bibinfo{volume}{60}},
  \bibinfo{pages}{3485} (\bibinfo{year}{1999}).

\bibitem[{\citenamefont{Solenov et~al.}(2009)\citenamefont{Solenov, Martin, and
  Mozyrsky}}]{solenov2009thermodynamical}
\bibinfo{author}{\bibfnamefont{D.}~\bibnamefont{Solenov}},
  \bibinfo{author}{\bibfnamefont{I.}~\bibnamefont{Martin}}, \bibnamefont{and}
  \bibinfo{author}{\bibfnamefont{D.}~\bibnamefont{Mozyrsky}},
  \bibinfo{journal}{Phys. Rev. B} \textbf{\bibinfo{volume}{79}},
  \bibinfo{pages}{132502} (\bibinfo{year}{2009}).

\bibitem[{\citenamefont{Kusunose et~al.}(2011)\citenamefont{Kusunose, Fuseya,
  and Miyake}}]{kusunose2011puzzle}
\bibinfo{author}{\bibfnamefont{H.}~\bibnamefont{Kusunose}},
  \bibinfo{author}{\bibfnamefont{Y.}~\bibnamefont{Fuseya}}, \bibnamefont{and}
  \bibinfo{author}{\bibfnamefont{K.}~\bibnamefont{Miyake}},
  \bibinfo{journal}{J. Phys. Soc. Jpn} \textbf{\bibinfo{volume}{80}},
  \bibinfo{pages}{054702} (\bibinfo{year}{2011}).

\bibitem[{\citenamefont{Fominov et~al.}(2015)\citenamefont{Fominov, Tanaka,
  Asano, and Eschrig}}]{FominovPRB2015}
\bibinfo{author}{\bibfnamefont{Y.~V.} \bibnamefont{Fominov}},
  \bibinfo{author}{\bibfnamefont{Y.}~\bibnamefont{Tanaka}},
  \bibinfo{author}{\bibfnamefont{Y.}~\bibnamefont{Asano}}, \bibnamefont{and}
  \bibinfo{author}{\bibfnamefont{M.}~\bibnamefont{Eschrig}},
  \bibinfo{journal}{Phys. Rev. B} \textbf{\bibinfo{volume}{91}},
  \bibinfo{pages}{144514} (\bibinfo{year}{2015}).

\bibitem[{\citenamefont{Bergeret et~al.}(2001)\citenamefont{Bergeret, Volkov,
  and Efetov}}]{BergeretPRL2001}
\bibinfo{author}{\bibfnamefont{F.~S.} \bibnamefont{Bergeret}},
  \bibinfo{author}{\bibfnamefont{A.~F.} \bibnamefont{Volkov}},
  \bibnamefont{and} \bibinfo{author}{\bibfnamefont{K.~B.}
  \bibnamefont{Efetov}}, \bibinfo{journal}{Phys. Rev. Lett.}
  \textbf{\bibinfo{volume}{86}}, \bibinfo{pages}{4096} (\bibinfo{year}{2001}).

\bibitem[{\citenamefont{Bergeret et~al.}(2005)\citenamefont{Bergeret, Volkov,
  and Efetov}}]{bergeret2005odd}
\bibinfo{author}{\bibfnamefont{F.~S.} \bibnamefont{Bergeret}},
  \bibinfo{author}{\bibfnamefont{A.~F.} \bibnamefont{Volkov}},
  \bibnamefont{and} \bibinfo{author}{\bibfnamefont{K.~B.}
  \bibnamefont{Efetov}}, \bibinfo{journal}{Rev. Mod. Phys.}
  \textbf{\bibinfo{volume}{77}}, \bibinfo{pages}{1321} (\bibinfo{year}{2005}).

\bibitem[{\citenamefont{Yokoyama et~al.}(2007)\citenamefont{Yokoyama, Tanaka,
  and Golubov}}]{yokoyama2007manifestation}
\bibinfo{author}{\bibfnamefont{T.}~\bibnamefont{Yokoyama}},
  \bibinfo{author}{\bibfnamefont{Y.}~\bibnamefont{Tanaka}}, \bibnamefont{and}
  \bibinfo{author}{\bibfnamefont{A.~A.} \bibnamefont{Golubov}},
  \bibinfo{journal}{Phys. Rev. B} \textbf{\bibinfo{volume}{75}},
  \bibinfo{pages}{134510} (\bibinfo{year}{2007}).

\bibitem[{\citenamefont{Houzet}(2008)}]{houzet2008ferromagnetic}
\bibinfo{author}{\bibfnamefont{M.}~\bibnamefont{Houzet}},
  \bibinfo{journal}{Phys. Rev. Lett.} \textbf{\bibinfo{volume}{101}},
  \bibinfo{pages}{057009} (\bibinfo{year}{2008}).

\bibitem[{\citenamefont{Eschrig and L{\"o}fwander}(2008)}]{EschrigNat2008}
\bibinfo{author}{\bibfnamefont{M.}~\bibnamefont{Eschrig}} \bibnamefont{and}
  \bibinfo{author}{\bibfnamefont{T.}~\bibnamefont{L{\"o}fwander}},
  \bibinfo{journal}{Nature Phys.} \textbf{\bibinfo{volume}{4}},
  \bibinfo{pages}{138} (\bibinfo{year}{2008}).

\bibitem[{\citenamefont{Linder et~al.}(2008)\citenamefont{Linder, Yokoyama, and
  Sudb\o{}}}]{LinderPRB2008}
\bibinfo{author}{\bibfnamefont{J.}~\bibnamefont{Linder}},
  \bibinfo{author}{\bibfnamefont{T.}~\bibnamefont{Yokoyama}}, \bibnamefont{and}
  \bibinfo{author}{\bibfnamefont{A.}~\bibnamefont{Sudb\o{}}},
  \bibinfo{journal}{Phys. Rev. B} \textbf{\bibinfo{volume}{77}},
  \bibinfo{pages}{174514} (\bibinfo{year}{2008}).

\bibitem[{\citenamefont{Cr\'epin et~al.}(2015)\citenamefont{Cr\'epin, Burset,
  and Trauzettel}}]{crepin2015odd}
\bibinfo{author}{\bibfnamefont{F.~m.~c.} \bibnamefont{Cr\'epin}},
  \bibinfo{author}{\bibfnamefont{P.}~\bibnamefont{Burset}}, \bibnamefont{and}
  \bibinfo{author}{\bibfnamefont{B.}~\bibnamefont{Trauzettel}},
  \bibinfo{journal}{Phys. Rev. B} \textbf{\bibinfo{volume}{92}},
  \bibinfo{pages}{100507} (\bibinfo{year}{2015}).

\bibitem[{\citenamefont{Yokoyama}(2012)}]{YokoyamaPRB2012}
\bibinfo{author}{\bibfnamefont{T.}~\bibnamefont{Yokoyama}},
  \bibinfo{journal}{Phys. Rev. B} \textbf{\bibinfo{volume}{86}},
  \bibinfo{pages}{075410} (\bibinfo{year}{2012}).

\bibitem[{\citenamefont{Black-Schaffer and
  Balatsky}(2012)}]{Black-SchafferPRB2012}
\bibinfo{author}{\bibfnamefont{A.~M.} \bibnamefont{Black-Schaffer}}
  \bibnamefont{and} \bibinfo{author}{\bibfnamefont{A.~V.}
  \bibnamefont{Balatsky}}, \bibinfo{journal}{Phys. Rev. B}
  \textbf{\bibinfo{volume}{86}}, \bibinfo{pages}{144506}
  (\bibinfo{year}{2012}).

\bibitem[{\citenamefont{Black-Schaffer and
  Balatsky}(2013{\natexlab{a}})}]{Black-SchafferPRB2013}
\bibinfo{author}{\bibfnamefont{A.~M.} \bibnamefont{Black-Schaffer}}
  \bibnamefont{and} \bibinfo{author}{\bibfnamefont{A.~V.}
  \bibnamefont{Balatsky}}, \bibinfo{journal}{Phys. Rev. B}
  \textbf{\bibinfo{volume}{87}}, \bibinfo{pages}{220506}
  (\bibinfo{year}{2013}{\natexlab{a}}).

\bibitem[{\citenamefont{Triola et~al.}(2014)\citenamefont{Triola, Rossi, and
  Balatsky}}]{TriolaPRB2014}
\bibinfo{author}{\bibfnamefont{C.}~\bibnamefont{Triola}},
  \bibinfo{author}{\bibfnamefont{E.}~\bibnamefont{Rossi}}, \bibnamefont{and}
  \bibinfo{author}{\bibfnamefont{A.~V.} \bibnamefont{Balatsky}},
  \bibinfo{journal}{Phys. Rev. B} \textbf{\bibinfo{volume}{89}},
  \bibinfo{pages}{165309} (\bibinfo{year}{2014}).

\bibitem[{\citenamefont{Tanaka and Golubov}(2007)}]{tanaka2007theory}
\bibinfo{author}{\bibfnamefont{Y.}~\bibnamefont{Tanaka}} \bibnamefont{and}
  \bibinfo{author}{\bibfnamefont{A.~A.} \bibnamefont{Golubov}},
  \bibinfo{journal}{Phys. Rev. Lett.} \textbf{\bibinfo{volume}{98}},
  \bibinfo{pages}{037003} (\bibinfo{year}{2007}).

\bibitem[{\citenamefont{Tanaka et~al.}(2007)\citenamefont{Tanaka, Tanuma, and
  Golubov}}]{TanakaPRB2007}
\bibinfo{author}{\bibfnamefont{Y.}~\bibnamefont{Tanaka}},
  \bibinfo{author}{\bibfnamefont{Y.}~\bibnamefont{Tanuma}}, \bibnamefont{and}
  \bibinfo{author}{\bibfnamefont{A.~A.} \bibnamefont{Golubov}},
  \bibinfo{journal}{Phys. Rev. B} \textbf{\bibinfo{volume}{76}},
  \bibinfo{pages}{054522} (\bibinfo{year}{2007}).

\bibitem[{\citenamefont{Linder et~al.}(2009)\citenamefont{Linder, Yokoyama,
  Sudb\o{}, and Eschrig}}]{LinderPRL2009}
\bibinfo{author}{\bibfnamefont{J.}~\bibnamefont{Linder}},
  \bibinfo{author}{\bibfnamefont{T.}~\bibnamefont{Yokoyama}},
  \bibinfo{author}{\bibfnamefont{A.}~\bibnamefont{Sudb\o{}}}, \bibnamefont{and}
  \bibinfo{author}{\bibfnamefont{M.}~\bibnamefont{Eschrig}},
  \bibinfo{journal}{Phys. Rev. Lett.} \textbf{\bibinfo{volume}{102}},
  \bibinfo{pages}{107008} (\bibinfo{year}{2009}).

\bibitem[{\citenamefont{Linder et~al.}(2010)\citenamefont{Linder, Sudb\o{},
  Yokoyama, Grein, and Eschrig}}]{LinderPRB2010_2}
\bibinfo{author}{\bibfnamefont{J.}~\bibnamefont{Linder}},
  \bibinfo{author}{\bibfnamefont{A.}~\bibnamefont{Sudb\o{}}},
  \bibinfo{author}{\bibfnamefont{T.}~\bibnamefont{Yokoyama}},
  \bibinfo{author}{\bibfnamefont{R.}~\bibnamefont{Grein}}, \bibnamefont{and}
  \bibinfo{author}{\bibfnamefont{M.}~\bibnamefont{Eschrig}},
  \bibinfo{journal}{Phys. Rev. B} \textbf{\bibinfo{volume}{81}},
  \bibinfo{pages}{214504} (\bibinfo{year}{2010}).

\bibitem[{\citenamefont{Tanaka et~al.}(2012)\citenamefont{Tanaka, Sato, and
  Nagaosa}}]{TanakaJPSJ2012}
\bibinfo{author}{\bibfnamefont{Y.}~\bibnamefont{Tanaka}},
  \bibinfo{author}{\bibfnamefont{M.}~\bibnamefont{Sato}}, \bibnamefont{and}
  \bibinfo{author}{\bibfnamefont{N.}~\bibnamefont{Nagaosa}},
  \bibinfo{journal}{J. Phys. Soc. Jpn} \textbf{\bibinfo{volume}{81}},
  \bibinfo{pages}{011013} (\bibinfo{year}{2012}).

\bibitem[{\citenamefont{Parhizgar and
  Black-Schaffer}(2014)}]{parhizgar_2014_prb}
\bibinfo{author}{\bibfnamefont{F.}~\bibnamefont{Parhizgar}} \bibnamefont{and}
  \bibinfo{author}{\bibfnamefont{A.~M.} \bibnamefont{Black-Schaffer}},
  \bibinfo{journal}{Phys. Rev. B} \textbf{\bibinfo{volume}{90}},
  \bibinfo{pages}{184517} (\bibinfo{year}{2014}).

\bibitem[{\citenamefont{Triola et~al.}(2016)\citenamefont{Triola, Badiane,
  Balatsky, and Rossi}}]{triola2016prl}
\bibinfo{author}{\bibfnamefont{C.}~\bibnamefont{Triola}},
  \bibinfo{author}{\bibfnamefont{D.~M.} \bibnamefont{Badiane}},
  \bibinfo{author}{\bibfnamefont{A.~V.} \bibnamefont{Balatsky}},
  \bibnamefont{and} \bibinfo{author}{\bibfnamefont{E.}~\bibnamefont{Rossi}},
  \bibinfo{journal}{Phys. Rev. Lett.} \textbf{\bibinfo{volume}{116}},
  \bibinfo{pages}{257001} (\bibinfo{year}{2016}).

\bibitem[{\citenamefont{Triola and Balatsky}(2016)}]{triolaprb2016}
\bibinfo{author}{\bibfnamefont{C.}~\bibnamefont{Triola}} \bibnamefont{and}
  \bibinfo{author}{\bibfnamefont{A.~V.} \bibnamefont{Balatsky}},
  \bibinfo{journal}{Phys. Rev. B} \textbf{\bibinfo{volume}{94}},
  \bibinfo{pages}{094518} (\bibinfo{year}{2016}).

\bibitem[{\citenamefont{Di~Bernardo
  et~al.}(2015{\natexlab{a}})\citenamefont{Di~Bernardo, Diesch, Gu, Linder,
  Divitini, Ducati, Scheer, Blamire, and Robinson}}]{di2015signature}
\bibinfo{author}{\bibfnamefont{A.}~\bibnamefont{Di~Bernardo}},
  \bibinfo{author}{\bibfnamefont{S.}~\bibnamefont{Diesch}},
  \bibinfo{author}{\bibfnamefont{Y.}~\bibnamefont{Gu}},
  \bibinfo{author}{\bibfnamefont{J.}~\bibnamefont{Linder}},
  \bibinfo{author}{\bibfnamefont{G.}~\bibnamefont{Divitini}},
  \bibinfo{author}{\bibfnamefont{C.}~\bibnamefont{Ducati}},
  \bibinfo{author}{\bibfnamefont{E.}~\bibnamefont{Scheer}},
  \bibinfo{author}{\bibfnamefont{M.~G.} \bibnamefont{Blamire}},
  \bibnamefont{and} \bibinfo{author}{\bibfnamefont{J.~W.}
  \bibnamefont{Robinson}}, \bibinfo{journal}{Nature Commun.}
  \textbf{\bibinfo{volume}{6}}, \bibinfo{pages}{8053}
  (\bibinfo{year}{2015}{\natexlab{a}}).

\bibitem[{\citenamefont{Di~Bernardo
  et~al.}(2015{\natexlab{b}})\citenamefont{Di~Bernardo, Salman, Wang, Amado,
  Egilmez, Flokstra, Suter, Lee, Zhao, Prokscha et~al.}}]{di2015intrinsic}
\bibinfo{author}{\bibfnamefont{A.}~\bibnamefont{Di~Bernardo}},
  \bibinfo{author}{\bibfnamefont{Z.}~\bibnamefont{Salman}},
  \bibinfo{author}{\bibfnamefont{X.~L.} \bibnamefont{Wang}},
  \bibinfo{author}{\bibfnamefont{M.}~\bibnamefont{Amado}},
  \bibinfo{author}{\bibfnamefont{M.}~\bibnamefont{Egilmez}},
  \bibinfo{author}{\bibfnamefont{M.~G.} \bibnamefont{Flokstra}},
  \bibinfo{author}{\bibfnamefont{A.}~\bibnamefont{Suter}},
  \bibinfo{author}{\bibfnamefont{S.~L.} \bibnamefont{Lee}},
  \bibinfo{author}{\bibfnamefont{J.~H.} \bibnamefont{Zhao}},
  \bibinfo{author}{\bibfnamefont{T.}~\bibnamefont{Prokscha}},
  \bibnamefont{et~al.}, \bibinfo{journal}{Phys. Rev. X}
  \textbf{\bibinfo{volume}{5}}, \bibinfo{pages}{041021}
  (\bibinfo{year}{2015}{\natexlab{b}}).

\bibitem[{\citenamefont{Linder and Balatsky}(2017)}]{linder2017odd}
\bibinfo{author}{\bibfnamefont{J.}~\bibnamefont{Linder}} \bibnamefont{and}
  \bibinfo{author}{\bibfnamefont{A.~V.} \bibnamefont{Balatsky}},
  \bibinfo{journal}{arXiv preprint arXiv:1709.03986}  (\bibinfo{year}{2017}).

\bibitem[{\citenamefont{Black-Schaffer and
  Balatsky}(2013{\natexlab{b}})}]{black2013odd}
\bibinfo{author}{\bibfnamefont{A.~M.} \bibnamefont{Black-Schaffer}}
  \bibnamefont{and} \bibinfo{author}{\bibfnamefont{A.~V.}
  \bibnamefont{Balatsky}}, \bibinfo{journal}{Phys. Rev. B}
  \textbf{\bibinfo{volume}{88}}, \bibinfo{pages}{104514}
  (\bibinfo{year}{2013}{\natexlab{b}}).

\bibitem[{\citenamefont{Komendov\'a et~al.}(2015)\citenamefont{Komendov\'a,
  Balatsky, and Black-Schaffer}}]{komendova2015experimentally}
\bibinfo{author}{\bibfnamefont{L.}~\bibnamefont{Komendov\'a}},
  \bibinfo{author}{\bibfnamefont{A.~V.} \bibnamefont{Balatsky}},
  \bibnamefont{and} \bibinfo{author}{\bibfnamefont{A.~M.}
  \bibnamefont{Black-Schaffer}}, \bibinfo{journal}{Phys. Rev. B}
  \textbf{\bibinfo{volume}{92}}, \bibinfo{pages}{094517}
  (\bibinfo{year}{2015}).

\bibitem[{\citenamefont{Komendov\'a and
  Black-Schaffer}(2017)}]{komendova2017odd}
\bibinfo{author}{\bibfnamefont{L.}~\bibnamefont{Komendov\'a}} \bibnamefont{and}
  \bibinfo{author}{\bibfnamefont{A.~M.} \bibnamefont{Black-Schaffer}},
  \bibinfo{journal}{Phys. Rev. Lett.} \textbf{\bibinfo{volume}{119}},
  \bibinfo{pages}{087001} (\bibinfo{year}{2017}).

\bibitem[{\citenamefont{Triola and Balatsky}(2017)}]{triola_prb_2017}
\bibinfo{author}{\bibfnamefont{C.}~\bibnamefont{Triola}} \bibnamefont{and}
  \bibinfo{author}{\bibfnamefont{A.~V.} \bibnamefont{Balatsky}},
  \bibinfo{journal}{Phys. Rev. B} \textbf{\bibinfo{volume}{95}},
  \bibinfo{pages}{224518} (\bibinfo{year}{2017}).

\bibitem[{\citenamefont{Maeno et~al.}(1994)\citenamefont{Maeno, Hashimoto,
  Yoshida, Nishizaki, Fujita, Bednorz, and
  Lichtenberg}}]{maeno1994superconductivity}
\bibinfo{author}{\bibfnamefont{Y.}~\bibnamefont{Maeno}},
  \bibinfo{author}{\bibfnamefont{H.}~\bibnamefont{Hashimoto}},
  \bibinfo{author}{\bibfnamefont{K.}~\bibnamefont{Yoshida}},
  \bibinfo{author}{\bibfnamefont{S.}~\bibnamefont{Nishizaki}},
  \bibinfo{author}{\bibfnamefont{T.}~\bibnamefont{Fujita}},
  \bibinfo{author}{\bibfnamefont{J.}~\bibnamefont{Bednorz}}, \bibnamefont{and}
  \bibinfo{author}{\bibfnamefont{F.}~\bibnamefont{Lichtenberg}},
  \bibinfo{journal}{Nature} \textbf{\bibinfo{volume}{372}},
  \bibinfo{pages}{532} (\bibinfo{year}{1994}).

\bibitem[{\citenamefont{Maeno et~al.}(2012)\citenamefont{Maeno, Kittaka,
  Nomura, Yonezawa, and Ishida}}]{maeno2012}
\bibinfo{author}{\bibfnamefont{Y.}~\bibnamefont{Maeno}},
  \bibinfo{author}{\bibfnamefont{S.}~\bibnamefont{Kittaka}},
  \bibinfo{author}{\bibfnamefont{T.}~\bibnamefont{Nomura}},
  \bibinfo{author}{\bibfnamefont{S.}~\bibnamefont{Yonezawa}}, \bibnamefont{and}
  \bibinfo{author}{\bibfnamefont{K.}~\bibnamefont{Ishida}},
  \bibinfo{journal}{J. Phys. Soc. Jpn} \textbf{\bibinfo{volume}{81}},
  \bibinfo{pages}{011009} (\bibinfo{year}{2012}).

\bibitem[{\citenamefont{Hunte et~al.}(2008)\citenamefont{Hunte, Jaroszynski,
  Gurevich, Larbalestier, Jin, Sefat, McGuire, Sales, Christen, and
  Mandrus}}]{hunte2008two}
\bibinfo{author}{\bibfnamefont{F.}~\bibnamefont{Hunte}},
  \bibinfo{author}{\bibfnamefont{J.}~\bibnamefont{Jaroszynski}},
  \bibinfo{author}{\bibfnamefont{A.}~\bibnamefont{Gurevich}},
  \bibinfo{author}{\bibfnamefont{D.}~\bibnamefont{Larbalestier}},
  \bibinfo{author}{\bibfnamefont{R.}~\bibnamefont{Jin}},
  \bibinfo{author}{\bibfnamefont{A.}~\bibnamefont{Sefat}},
  \bibinfo{author}{\bibfnamefont{M.~A.} \bibnamefont{McGuire}},
  \bibinfo{author}{\bibfnamefont{B.~C.} \bibnamefont{Sales}},
  \bibinfo{author}{\bibfnamefont{D.~K.} \bibnamefont{Christen}},
  \bibnamefont{and} \bibinfo{author}{\bibfnamefont{D.}~\bibnamefont{Mandrus}},
  \bibinfo{journal}{Nature} \textbf{\bibinfo{volume}{453}},
  \bibinfo{pages}{903} (\bibinfo{year}{2008}).

\bibitem[{\citenamefont{Kamihara et~al.}(2008)\citenamefont{Kamihara, Watanabe,
  Hirano, and Hosono}}]{kamihara2008iron}
\bibinfo{author}{\bibfnamefont{Y.}~\bibnamefont{Kamihara}},
  \bibinfo{author}{\bibfnamefont{T.}~\bibnamefont{Watanabe}},
  \bibinfo{author}{\bibfnamefont{M.}~\bibnamefont{Hirano}}, \bibnamefont{and}
  \bibinfo{author}{\bibfnamefont{H.}~\bibnamefont{Hosono}},
  \bibinfo{journal}{J. Am. Chem. Soc.} \textbf{\bibinfo{volume}{130}},
  \bibinfo{pages}{3296} (\bibinfo{year}{2008}).

\bibitem[{\citenamefont{Ishida et~al.}(2009)\citenamefont{Ishida, Nakai, and
  Hosono}}]{ishida2009extent}
\bibinfo{author}{\bibfnamefont{K.}~\bibnamefont{Ishida}},
  \bibinfo{author}{\bibfnamefont{Y.}~\bibnamefont{Nakai}}, \bibnamefont{and}
  \bibinfo{author}{\bibfnamefont{H.}~\bibnamefont{Hosono}},
  \bibinfo{journal}{J. Phys. Soc. Jpn} \textbf{\bibinfo{volume}{78}},
  \bibinfo{pages}{062001} (\bibinfo{year}{2009}).

\bibitem[{\citenamefont{Cvetkovic and
  Tesanovic}(2009)}]{cvetkovic2009multiband}
\bibinfo{author}{\bibfnamefont{V.}~\bibnamefont{Cvetkovic}} \bibnamefont{and}
  \bibinfo{author}{\bibfnamefont{Z.}~\bibnamefont{Tesanovic}},
  \bibinfo{journal}{EPL} \textbf{\bibinfo{volume}{85}}, \bibinfo{pages}{37002}
  (\bibinfo{year}{2009}).

\bibitem[{\citenamefont{Stewart}(2011)}]{stewart2011superconductivity}
\bibinfo{author}{\bibfnamefont{G.}~\bibnamefont{Stewart}},
  \bibinfo{journal}{Reviews of Modern Physics} \textbf{\bibinfo{volume}{83}},
  \bibinfo{pages}{1589} (\bibinfo{year}{2011}).

\bibitem[{\citenamefont{Nagamatsu et~al.}(2001)\citenamefont{Nagamatsu,
  Nakagawa, Muranaka, Zenitani, and Akimitsu}}]{nagamatsu2001superconductivity}
\bibinfo{author}{\bibfnamefont{J.}~\bibnamefont{Nagamatsu}},
  \bibinfo{author}{\bibfnamefont{N.}~\bibnamefont{Nakagawa}},
  \bibinfo{author}{\bibfnamefont{T.}~\bibnamefont{Muranaka}},
  \bibinfo{author}{\bibfnamefont{Y.}~\bibnamefont{Zenitani}}, \bibnamefont{and}
  \bibinfo{author}{\bibfnamefont{J.}~\bibnamefont{Akimitsu}},
  \bibinfo{journal}{nature} \textbf{\bibinfo{volume}{410}}, \bibinfo{pages}{63}
  (\bibinfo{year}{2001}).

\bibitem[{\citenamefont{Bouquet et~al.}(2001)\citenamefont{Bouquet, Fisher,
  Phillips, Hinks, and Jorgensen}}]{bouquet2001specific}
\bibinfo{author}{\bibfnamefont{F.}~\bibnamefont{Bouquet}},
  \bibinfo{author}{\bibfnamefont{R.}~\bibnamefont{Fisher}},
  \bibinfo{author}{\bibfnamefont{N.}~\bibnamefont{Phillips}},
  \bibinfo{author}{\bibfnamefont{D.}~\bibnamefont{Hinks}}, \bibnamefont{and}
  \bibinfo{author}{\bibfnamefont{J.}~\bibnamefont{Jorgensen}},
  \bibinfo{journal}{Phys. Rev. Lett.} \textbf{\bibinfo{volume}{87}},
  \bibinfo{pages}{047001} (\bibinfo{year}{2001}).

\bibitem[{\citenamefont{Brinkman et~al.}(2002)\citenamefont{Brinkman, Golubov,
  Rogalla, Dolgov, Kortus, Kong, Jepsen, and Andersen}}]{brinkman2002multiband}
\bibinfo{author}{\bibfnamefont{A.}~\bibnamefont{Brinkman}},
  \bibinfo{author}{\bibfnamefont{A.}~\bibnamefont{Golubov}},
  \bibinfo{author}{\bibfnamefont{H.}~\bibnamefont{Rogalla}},
  \bibinfo{author}{\bibfnamefont{O.}~\bibnamefont{Dolgov}},
  \bibinfo{author}{\bibfnamefont{J.}~\bibnamefont{Kortus}},
  \bibinfo{author}{\bibfnamefont{Y.}~\bibnamefont{Kong}},
  \bibinfo{author}{\bibfnamefont{O.}~\bibnamefont{Jepsen}}, \bibnamefont{and}
  \bibinfo{author}{\bibfnamefont{O.}~\bibnamefont{Andersen}},
  \bibinfo{journal}{Phys. Rev. B} \textbf{\bibinfo{volume}{65}},
  \bibinfo{pages}{180517} (\bibinfo{year}{2002}).

\bibitem[{\citenamefont{Golubov et~al.}(2002)\citenamefont{Golubov, Kortus,
  Dolgov, Jepsen, Kong, Andersen, Gibson, Ahn, and
  Kremer}}]{golubov2002specific}
\bibinfo{author}{\bibfnamefont{A.}~\bibnamefont{Golubov}},
  \bibinfo{author}{\bibfnamefont{J.}~\bibnamefont{Kortus}},
  \bibinfo{author}{\bibfnamefont{O.}~\bibnamefont{Dolgov}},
  \bibinfo{author}{\bibfnamefont{O.}~\bibnamefont{Jepsen}},
  \bibinfo{author}{\bibfnamefont{Y.}~\bibnamefont{Kong}},
  \bibinfo{author}{\bibfnamefont{O.}~\bibnamefont{Andersen}},
  \bibinfo{author}{\bibfnamefont{B.}~\bibnamefont{Gibson}},
  \bibinfo{author}{\bibfnamefont{K.}~\bibnamefont{Ahn}}, \bibnamefont{and}
  \bibinfo{author}{\bibfnamefont{R.}~\bibnamefont{Kremer}},
  \bibinfo{journal}{J. Phys.: Condens. Matter} \textbf{\bibinfo{volume}{14}},
  \bibinfo{pages}{1353} (\bibinfo{year}{2002}).

\bibitem[{\citenamefont{Iavarone et~al.}(2002)\citenamefont{Iavarone,
  Karapetrov, Koshelev, Kwok, Crabtree, Hinks, Kang, Choi, Kim, Kim
  et~al.}}]{iavarone2002two}
\bibinfo{author}{\bibfnamefont{M.}~\bibnamefont{Iavarone}},
  \bibinfo{author}{\bibfnamefont{G.}~\bibnamefont{Karapetrov}},
  \bibinfo{author}{\bibfnamefont{A.}~\bibnamefont{Koshelev}},
  \bibinfo{author}{\bibfnamefont{W.}~\bibnamefont{Kwok}},
  \bibinfo{author}{\bibfnamefont{G.}~\bibnamefont{Crabtree}},
  \bibinfo{author}{\bibfnamefont{D.}~\bibnamefont{Hinks}},
  \bibinfo{author}{\bibfnamefont{W.}~\bibnamefont{Kang}},
  \bibinfo{author}{\bibfnamefont{E.-M.} \bibnamefont{Choi}},
  \bibinfo{author}{\bibfnamefont{H.~J.} \bibnamefont{Kim}},
  \bibinfo{author}{\bibfnamefont{H.-J.} \bibnamefont{Kim}},
  \bibnamefont{et~al.}, \bibinfo{journal}{Phys. Rev. Lett.}
  \textbf{\bibinfo{volume}{89}}, \bibinfo{pages}{187002}
  (\bibinfo{year}{2002}).

\bibitem[{\citenamefont{Xia et~al.}(2006)\citenamefont{Xia, Maeno, Beyersdorf,
  Fejer, and Kapitulnik}}]{xia_prl_2006}
\bibinfo{author}{\bibfnamefont{J.}~\bibnamefont{Xia}},
  \bibinfo{author}{\bibfnamefont{Y.}~\bibnamefont{Maeno}},
  \bibinfo{author}{\bibfnamefont{P.~T.} \bibnamefont{Beyersdorf}},
  \bibinfo{author}{\bibfnamefont{M.~M.} \bibnamefont{Fejer}}, \bibnamefont{and}
  \bibinfo{author}{\bibfnamefont{A.}~\bibnamefont{Kapitulnik}},
  \bibinfo{journal}{Phys. Rev. Lett.} \textbf{\bibinfo{volume}{97}},
  \bibinfo{pages}{167002} (\bibinfo{year}{2006}).

\bibitem[{\citenamefont{Wysoki\ifmmode~\acute{n}\else \'{n}\fi{}ski
  et~al.}(2012)\citenamefont{Wysoki\ifmmode~\acute{n}\else \'{n}\fi{}ski,
  Annett, and Gy\"orffy}}]{wysokinski_2012_prl}
\bibinfo{author}{\bibfnamefont{K.~I.}
  \bibnamefont{Wysoki\ifmmode~\acute{n}\else \'{n}\fi{}ski}},
  \bibinfo{author}{\bibfnamefont{J.~F.} \bibnamefont{Annett}},
  \bibnamefont{and} \bibinfo{author}{\bibfnamefont{B.~L.}
  \bibnamefont{Gy\"orffy}}, \bibinfo{journal}{Phys. Rev. Lett.}
  \textbf{\bibinfo{volume}{108}}, \bibinfo{pages}{077004}
  (\bibinfo{year}{2012}).

\bibitem[{\citenamefont{Taylor and Kallin}(2012)}]{taylor_prl_2012}
\bibinfo{author}{\bibfnamefont{E.}~\bibnamefont{Taylor}} \bibnamefont{and}
  \bibinfo{author}{\bibfnamefont{C.}~\bibnamefont{Kallin}},
  \bibinfo{journal}{Phys. Rev. Lett.} \textbf{\bibinfo{volume}{108}},
  \bibinfo{pages}{157001} (\bibinfo{year}{2012}).

\bibitem[{\citenamefont{Gradhand et~al.}(2013)\citenamefont{Gradhand,
  Wysokinski, Annett, and Gy\"orffy}}]{gradhand_2013_prb}
\bibinfo{author}{\bibfnamefont{M.}~\bibnamefont{Gradhand}},
  \bibinfo{author}{\bibfnamefont{K.~I.} \bibnamefont{Wysokinski}},
  \bibinfo{author}{\bibfnamefont{J.~F.} \bibnamefont{Annett}},
  \bibnamefont{and} \bibinfo{author}{\bibfnamefont{B.~L.}
  \bibnamefont{Gy\"orffy}}, \bibinfo{journal}{Phys. Rev. B}
  \textbf{\bibinfo{volume}{88}}, \bibinfo{pages}{094504}
  (\bibinfo{year}{2013}).

\bibitem[{\citenamefont{Lutchyn et~al.}(2009)\citenamefont{Lutchyn, Nagornykh,
  and Yakovenko}}]{lutchyn_prb_2009}
\bibinfo{author}{\bibfnamefont{R.~M.} \bibnamefont{Lutchyn}},
  \bibinfo{author}{\bibfnamefont{P.}~\bibnamefont{Nagornykh}},
  \bibnamefont{and} \bibinfo{author}{\bibfnamefont{V.~M.}
  \bibnamefont{Yakovenko}}, \bibinfo{journal}{Phys. Rev. B}
  \textbf{\bibinfo{volume}{80}}, \bibinfo{pages}{104508}
  (\bibinfo{year}{2009}).

\bibitem[{\citenamefont{Joynt and Taillefer}(2002)}]{joynt_rmp_2002}
\bibinfo{author}{\bibfnamefont{R.}~\bibnamefont{Joynt}} \bibnamefont{and}
  \bibinfo{author}{\bibfnamefont{L.}~\bibnamefont{Taillefer}},
  \bibinfo{journal}{Rev. Mod. Phys.} \textbf{\bibinfo{volume}{74}},
  \bibinfo{pages}{235} (\bibinfo{year}{2002}).

\bibitem[{\citenamefont{Taillefer et~al.}(1987)\citenamefont{Taillefer,
  Newbury, Lonzarich, Fisk, and Smith}}]{taillefer_1987}
\bibinfo{author}{\bibfnamefont{L.}~\bibnamefont{Taillefer}},
  \bibinfo{author}{\bibfnamefont{R.}~\bibnamefont{Newbury}},
  \bibinfo{author}{\bibfnamefont{G.}~\bibnamefont{Lonzarich}},
  \bibinfo{author}{\bibfnamefont{Z.}~\bibnamefont{Fisk}}, \bibnamefont{and}
  \bibinfo{author}{\bibfnamefont{J.}~\bibnamefont{Smith}}, \bibinfo{journal}{J.
  Magn. Magn. Mater.} \textbf{\bibinfo{volume}{63-64}}, \bibinfo{pages}{372 }
  (\bibinfo{year}{1987}).

\bibitem[{\citenamefont{Taillefer and Lonzarich}(1988)}]{taillefer_prl_1988}
\bibinfo{author}{\bibfnamefont{L.}~\bibnamefont{Taillefer}} \bibnamefont{and}
  \bibinfo{author}{\bibfnamefont{G.~G.} \bibnamefont{Lonzarich}},
  \bibinfo{journal}{Phys. Rev. Lett.} \textbf{\bibinfo{volume}{60}},
  \bibinfo{pages}{1570} (\bibinfo{year}{1988}).

\bibitem[{\citenamefont{Wang et~al.}(1987)\citenamefont{Wang, Norman, Albers,
  Boring, Pickett, Krakauer, and Christensen}}]{wang_prb_1987}
\bibinfo{author}{\bibfnamefont{C.~S.} \bibnamefont{Wang}},
  \bibinfo{author}{\bibfnamefont{M.~R.} \bibnamefont{Norman}},
  \bibinfo{author}{\bibfnamefont{R.~C.} \bibnamefont{Albers}},
  \bibinfo{author}{\bibfnamefont{A.~M.} \bibnamefont{Boring}},
  \bibinfo{author}{\bibfnamefont{W.~E.} \bibnamefont{Pickett}},
  \bibinfo{author}{\bibfnamefont{H.}~\bibnamefont{Krakauer}}, \bibnamefont{and}
  \bibinfo{author}{\bibfnamefont{N.~E.} \bibnamefont{Christensen}},
  \bibinfo{journal}{Phys. Rev. B} \textbf{\bibinfo{volume}{35}},
  \bibinfo{pages}{7260} (\bibinfo{year}{1987}).

\bibitem[{\citenamefont{Norman et~al.}(1988)\citenamefont{Norman, Albers,
  Boring, and Christensen}}]{norman_ssc_1988}
\bibinfo{author}{\bibfnamefont{M.}~\bibnamefont{Norman}},
  \bibinfo{author}{\bibfnamefont{R.}~\bibnamefont{Albers}},
  \bibinfo{author}{\bibfnamefont{A.}~\bibnamefont{Boring}}, \bibnamefont{and}
  \bibinfo{author}{\bibfnamefont{N.}~\bibnamefont{Christensen}},
  \bibinfo{journal}{Solid State Commun.} \textbf{\bibinfo{volume}{68}},
  \bibinfo{pages}{245 } (\bibinfo{year}{1988}), ISSN \bibinfo{issn}{0038-1098}.

\bibitem[{\citenamefont{McMullan et~al.}(2008)\citenamefont{McMullan, Norman,
  Huxley, Doiron-Leyraud, Flouquet, Lonzarich, McCollam, and
  Julian}}]{mcmullan2008fermi}
\bibinfo{author}{\bibfnamefont{G.}~\bibnamefont{McMullan}},
  \bibinfo{author}{\bibfnamefont{M.}~\bibnamefont{Norman}},
  \bibinfo{author}{\bibfnamefont{A.}~\bibnamefont{Huxley}},
  \bibinfo{author}{\bibfnamefont{N.}~\bibnamefont{Doiron-Leyraud}},
  \bibinfo{author}{\bibfnamefont{J.}~\bibnamefont{Flouquet}},
  \bibinfo{author}{\bibfnamefont{G.}~\bibnamefont{Lonzarich}},
  \bibinfo{author}{\bibfnamefont{A.}~\bibnamefont{McCollam}}, \bibnamefont{and}
  \bibinfo{author}{\bibfnamefont{S.}~\bibnamefont{Julian}},
  \bibinfo{journal}{New J. Phys.} \textbf{\bibinfo{volume}{10}},
  \bibinfo{pages}{053029} (\bibinfo{year}{2008}).

\bibitem[{\citenamefont{Heal and Williams}(1955)}]{heal_1955}
\bibinfo{author}{\bibfnamefont{T.~J.} \bibnamefont{Heal}} \bibnamefont{and}
  \bibinfo{author}{\bibfnamefont{G.~I.} \bibnamefont{Williams}},
  \bibinfo{journal}{Acta Crystallographica} \textbf{\bibinfo{volume}{8}},
  \bibinfo{pages}{494} (\bibinfo{year}{1955}).

\bibitem[{\citenamefont{Walko et~al.}(2001)\citenamefont{Walko, Hong,
  Chandrasekhar~Rao, Wawrzak, Seidman, Halperin, and Bedzyk}}]{walko_2001_prb}
\bibinfo{author}{\bibfnamefont{D.~A.} \bibnamefont{Walko}},
  \bibinfo{author}{\bibfnamefont{J.-I.} \bibnamefont{Hong}},
  \bibinfo{author}{\bibfnamefont{T.~V.} \bibnamefont{Chandrasekhar~Rao}},
  \bibinfo{author}{\bibfnamefont{Z.}~\bibnamefont{Wawrzak}},
  \bibinfo{author}{\bibfnamefont{D.~N.} \bibnamefont{Seidman}},
  \bibinfo{author}{\bibfnamefont{W.~P.} \bibnamefont{Halperin}},
  \bibnamefont{and} \bibinfo{author}{\bibfnamefont{M.~J.}
  \bibnamefont{Bedzyk}}, \bibinfo{journal}{Phys. Rev. B}
  \textbf{\bibinfo{volume}{63}}, \bibinfo{pages}{054522}
  (\bibinfo{year}{2001}).

\end{thebibliography}

\end{document}